\documentstyle[prb,twocolumn,aps]{revtex}
\newcommand{\be}{\begin{eqnarray}}
\newcommand{\ee}{\end{eqnarray}}
\newcommand{\OP}{\psi}

\newcommand{\F}{{\cal F}}

\begin{document}

\title{Perturbation Expansion in Phase-Ordering Kinetics: I.  Scalar Order
Parameter}
\author{Gene F. Mazenko }
\address{The James Franck Institute and the Department of Physics\\
The University of Chicago\\
Chicago, Illinois 60637}
\date{\today}
\maketitle
%

%
%
\begin{abstract}

A consistent perturbation theory expansion is presented for
phase-ordering kinetics in the case of a nonconserved scalar order
parameter.  At zeroth order in this expansion one obtains the
theory due to Ohta,  Jasnow and Kawasaki (OJK).  At the next nontrivial
order in the expansion,
worked out in d dimensions, one has small corrections
to the OJK result for
the nonequilibrium exponent $\lambda$ and the introduction of a new
exponent $\nu$ governing the algebraic component of the decay of the
order parameter scaling function at large scaled distances.

\end{abstract}


\section{Introduction}

Significant progress has been made on the theory of phase-ordering
kinetics\cite{2} using methods that introduce auxiliary fields that are
taken to have gaussian
statistics.  These theories well describe the qualitative scaling
features of ordering in unstable systems.
The methods developed in
OJK\cite{OJK} and TUG\cite{TUG} each have appealing aspects and
separately give
good descriptions of different aspects
of the ordering problem.
A major lingering
question is why these methods work as well as they do and how they can
be reconciled and improved.  Thus there has been a
search\cite{BH,post,postn}
for the field
theory description where OJK or TUG is the zeroth-order approximation
in some {\bf systematic} expansion.  Such an expansion is presented in
this paper\cite{pre}.  It has the OJK result as its zeroth-order approximation.
More importantly
it indicates how one goes forward to improve on these theories.
The case of a scalar order parameter is treated in this paper.
These ideas are generalized in a companion paper to the case of
the $n$-vector model and systems with continuous symmetry in the
disordered state.

The problem of interest is the restoration of equilibrium in a
system rendered unstable by a rapid temperature quench to a
regime where the final state corresponds to a broken discrete
symmetry.  The ordering is controlled in this case by the
decreasing area of domain walls separating the competing final
degenerate states.
The two coexisting theories, OJK and TUG, have
been useful in understanding certain aspects of this problem.
The theory developed in TUG has led to nontrivial expressions for the
nonequilibrium exponent $\lambda$, defined in detail below, which are
in good agreement with values known exactly or from simulation data.
However, as discussed by
Mazenko and Wickham\cite{fluc}, in the TUG approach
the auxiliary-field correlation function exhibits
a nonanalytic structure at short-scaled distance
which
leads to unphysical results
when used in calculations of defect
densitites\cite{LM92,MW97}
and defect velocity
distributions\cite{Maz97,vvv}
for systems with continuous symmetry in the disordered
states.  More recently they showed\cite{OJKpap} that the OJK
theory can be derived from the
exact continuity equation for the defect densities for point and line
defects and leads to smooth physical results for defect properties.
On the other hand, the OJK result is only compatible with
rather trivial results for the exponent $\lambda$.
Thus, at this point, one does not have a theory which is both
smooth enough for treating defect dynamics and yet robust enough
to give nontrivial results for the nonequilibrium exponent
$\lambda$.

In principle the task for the theorist in this problem
may seem obvious:  Linearize
the order parameter equation of motion and formally
arrive at a gaussian field theory.   Then do perturbation theory in
the remaining nonlinearity.  This is the well known path in
conventional field theory.  The problem, however, becomes clear when one
looks at the simplest form for the equation of motion satisfied by the
order parameter $\psi$ in dimensionless units
\be
\frac{\partial \psi}{\partial t}= \psi -\psi^{3}+\nabla^{2}\psi ~~~.
\label{eq:1}
\ee
It is clear from this form that there is no dimensionless
nonlinear coupling
in which to expand.  If one expands in the nonlinear term
$\psi^{3}$ one obtains exponential growth in time (the
Cahn-Hilliard theory\cite{CH} in the case where one has a conserved order
parameter) which, as a zeroth-order approximation, does not
include any of the basic qualitative features of the long-time
ordering.  The reason is that the nonlinearity is essential in
stabilizing the growth in the long-time limit.  Indeed the
combination $\psi -\psi^{3}$ must become small as the system
orders.  A mean field theory
by, Langer, Bar-on, and Miller\cite{Langer}, is an improvement but
is ultimately flawed by the
inability to treat the separation of the two characteristic
lengths in the problem.  The dominant scaling length, $L(t)$,
 grows algebraically with time and
characterizes the average separation between defects in the system.
The other length is the equilibrium correlation length $\xi_{E}$.
Clearly at sufficiently long times $L(t)\gg\xi_{E}$.  This two length
problem was addressed by Mazenko, Valls and Zannetti\cite{MVZ}
(MVZ) who
argued that the solution to this problem is to separate the order
parameter field $\psi$ into a peak contribution $\sigma$ and
a fluctuating contribution $u$,
$
\psi = \sigma +u ~.
$
Then $\sigma$ is associated with ordering on the length scale $L(t)$
and $u$ with equilibration on the length scale $\xi_{E}$.  In the
structure factor the $\sigma$ variable is identified with the
growth of a Bragg peak with width $L^{-1}(t)$ and $u$ is identified with
the Ornstein-Zernike contribution with width governed by
$\xi_{E}^{-1}$.  One reason for going over to the auxiliary field method is
to insure that the Bragg peak grows with the proper weight which
corresponds to the equilibrium average
of the
order parameter squared.

The work of OJK predates that of MVZ but fits into the picture developed
there. OJK
ignore the fluctuations ($u=0$) and assume $\sigma$ is a
function of a gaussian auxiliary field which is diffusive in
nature.  There have been a number of subsequent
papers\cite{OP,YJ,PR} clarifying
the nature of the OJK result.  More recently, in the work of
Bray and Humayun\cite{BH}, there has been an
effort
to derive the OJK theory in a systematic fashion.  This approach will
be discussed below in section III.

An alternative implementation of the ideas of MVZ was carried out in
TUG.  In this case, as discussed below, the mapping of
the order parameter onto an auxiliary field $m$
is motivated by the idea
that $m$ measures the distance to the nearest defect.  Coupled with the
assumption that the  auxiliary field is gaussian, the equation of
motion satisfied by
$\sigma$ is enforced, and one obtains a theory describing most of the features
satisfied by the order parameter scaling function.

All of these
theories with auxiliary fields satisifying gaussian statistics have
been lumped together as
{\it gaussian ~closure~approximations}\cite{YOS}.
It has been clear for some time that we have needed to have theories which
go further\cite{BBS,BB}.
Previous efforts at
a post-gaussian approximations\cite{post,postn} had some success but are
difficult to control.

How does one construct a perturbation theory expansion for the
auxiliary field $m$?  This theory will be unusual since, on average,
$m$ is growing with time.  Thus standard polynomial nonlinearities in
$m$ would be a problem.  As discussed in section III there are some
difficult technical problems in developing on expansion method which is
self-consistent.  The resolution to this problem is to first choose
the proper introduction of the auxiliary field $m$, and then organize
the treatment of the associated nonstandard nonlinear field theory
through the introdution of an expansion parameter, $\phi_{p}$.
At zeroth
order in this parameter one obtains the theory due to OJK and at
second order one finds expressions for the corrections to the
nontrivial exponents characterizing the ordering in these systems.
This parameter, $\phi_{p}$,
is associated with
the unusual nature of the nonlinearity in the
problem.  Instead of a polynomial nonlinearity one has the
{\it sign} of the field giving the driving nonlinear terms in
the equation of motion for the auxiliary field.
This
expansion in $\phi_{p}$ appears well behaved
order by order in perturbation theory and in some ways
is similar to
the bare expansion in the
quartic coupling in $\phi^{4}$ field theory.  One has, for example,
the possibility  of
resummation or use of renormalization group methods.  We
appear to be fortunate in this case since lower-order
approximations
appear to work well, give reasonable results for anamolous
dimensions, without the need for extensive reorganization.
One encounters a rather straightforward exponentiation of
logarithmic
divergences.  Unlike in critical phenomena the logs appear for
all dimensionality $d$.  In this case they are driven by internal
time integrations.

A key assertion in the theories developed previously is that the
auxiliary field can be treated as having gaussian statistics.  It
should be understood
that a theory with this feature at zeroth order must have the
property that
all of the
higher order cumulants for the field $m$ must vanish at this order.
It is shown
here that indeed the $n$-point
cumulant is of ${\cal O}(\frac{n}{2}-1)$
in the expansion parameter.  Thus the two-point cumulant, $G_{2}$,
which enters the determination of the order-parameter correlation function
at lowest order, is of ${\cal O}(0)$  as expected.  The four-point
cumulant is of ${\cal O}(1)$ and so on.  This has the consequence
that any function of the auxiliary field $m$  can be expressed
in terms of these cumulants and evaluated in perturbation theory.

Turning to the question of the existence of a small parameter, it
appears that the most direct connection of the expansion in
$\phi_{p}$ to a more conventional expansion parameter is to the
$1/n$ expansion in the $n-$vector model.  This will be discussed in
detail in paper II in this series.  This expansion does not appear
to be related to a $1/d$ expansion\cite{1/d}.
Thus the method developed here has more in common with methods in
critical phenomena where one works in the spatial dimension of interest
as compared, for example, to the $\epsilon$-expansion about four
dimensions.

\section{Overview}

\subsection{Setting Up the Problem}

The system studied here is the domain-wall dynamics generated by the
time-dependent Ginzburg-Landau (TDGL)
model satisfied by a nonconserved scalar order
parameter $\OP (\vec{r},t)$:
\be
\frac{\partial \psi}{\partial t}=
-\Gamma \frac{\delta F}{\delta \psi}  +\eta =K[\OP ]
\label{eq:2}
\ee
where $\Gamma $ is a kinetic coefficient,  $F$ is a Ginzburg-Landau
effective free energy assumed to be of the form
\be
F=\int ~d^{d}r \biggl( \frac{c}{2}(\nabla \OP)^{2}
+V(\OP )\biggr)
\ee
where $c > 0$ and the potential $V$ is assumed to be of the
symmetric
degenerate double-well
form. We expect only these general properties
of $V$ will be important in what follows.
$\eta $ is a thermal noise which is related to $\Gamma$
by a
fluctuation-dissipation theorem.
We assume that the quench is from a high temperature ($T_{I}>T_{c}$),
where the system is disordered, to zero temperature where the noise
can be set to zero ($\eta =0$).
It is believed\cite{2} that our final results are independent of
the exact nature of
the initial state, provided it is a disordered state with short-ranged
correlations.

If we rescale lengths and times we can put our equation of motion in
the dimensionless form
\be
\Lambda (1) \psi (1)=-V'[\OP (1)]
\label{eq:5}
\ee
where the diffusion operator
\be
\Lambda (1)=\frac{\partial}{\partial t_{1}}-\nabla^{2}_{1}
\ee
is introduced along with the short-hand notation that $1$ denotes
$({\bf r_{1}},t_{1})$.  Eq.(\ref{eq:1}) is just the special case of
the potential $V=-\frac{1}{2}\psi^{2}+\frac{1}{4}\psi^{4}$.

\subsection{Summary}

It is well
established\cite{2}
that for late times following a quench from the disordered to the
ordered phase the dynamics obey scaling and the system can be described in
terms of a single growing length, $L(t)$, which is characteristic of the
spacing between defects. In this scaling regime the order-parameter
correlation function has a universal scaling form

\be
\label{EQ:OPCOR}
C(12) \equiv \langle \OP (1) \OP (2) \rangle= \psi_{0}^{2} \F(x,t_{1}/t_{2})
\ee
where $\psi_{0}$ is the magnitude of the order-parameter in the
ordered phase.
The scaled length $x$ is defined as
$\vec{x} = (\vec{r}_{1}-\vec{r}_{2})/L(T)$ where, for the
non-conserved order parameter case considered here, the growth law goes as
$L(T) \sim T^{1/2}$ where
$T=\frac{1}{2}(t_{1}+t_{2})$.
In the case of the autocorrelation function
$\vec{r}_{1}=\vec{r}_{2}=\vec{r}$ we have\cite{FH,LM91}
\be
\langle \OP (\vec{r},t_{1}) \OP (\vec{r},t_{2}) \rangle
\approx \left(\frac{\sqrt{t_{1}t_{2}}}{T}\right)^{\lambda}
\ee
where $\lambda$ is a nontrivial nonequilibrium exponent and
either $t_{1}$ or $t_{2}$ is much larger than the other.
At equal-times and short-scaled distances
\be
\F(x)\equiv \F(x,1) = 1 -\alpha |x|+\cdots ~~~.
\label{eq:56}
\ee
This nonanalytic behavior as a function of $x$ is indicative of
Porod's law\cite{porod} as conventionally given in
terms of the Fourier transform
\be
{\cal F}(Q)\approx Q^{-(1+d)}
\ee
for large scaled wavenumber $Q$.  That all of the higher order terms in
Eq.(\ref{eq:56}) are odd in $|x|$ is known as the
Tomita sum rule\cite{tomita}.  It appears, as we will discuss
below, that the large $x$ behavior can, with proper definition of
$x$, be put in the form
\be
\F (x) \approx \frac{1}{x^{\nu}}e^{-\frac{1}{2}x^{2}}
\ee
where $\nu$ is a nontrivial subdominant exponent\cite{TUG1}.

The main results determined in this paper is an explicit
determination of
$\F (x,t_{1}/t_{2})$ in perturbation theory.  At zeroth order
we obtain the OJK result
\be
\F (x,t_{1}/t_{2})=\frac{2}{\pi}sin^{-1}\left[\Phi_{0}(t_{1},t_{2})
e^{-\frac{1}{2}x^{2}}\right]
\ee
where
\be
\Phi_{0}(t_{1},t_{2})=\left(\frac{\sqrt{t_{1}t_{2}}}{T}\right)
^{\lambda_{0}}
\ee
where $\lambda_{0}=d/2$ and $\nu_{0}=0$.  Going to next order,
${\cal O}(1)$,
in the
expansion, discussed in detail in section X, we find no change in the
indices $\lambda$ and $\nu$ but quantitative changes in
$\F(x,t_{1}/t_{2})$.  At ${\cal O}(2)$ both $\lambda$ and $\nu$ are
shifted.
The exponent $\lambda$ is given  at ${\cal O}(2)$ by
\be
\lambda =\frac{d}{2}+\omega^{2}\frac{2^{d}M_{d}}{3^{d/2+1}}
\label{eq:14}
\ee
where the  dimensionality dependent quantities, $\omega $, $K_{d}$ and
$M_{d}$,  are determined by
\be
2\omega+\omega^{2}2^{d}\left(K_{d}+
\frac{M_{d}}{3^{d/2+1}}\right)=1+\frac{d}{2}
\label{eq:6}
\ee
\be
K_{d}=\int_{0}^{1}dz \frac{z^{d/2-1}}{[(1+z)(3-z)]^{d/2}}
\label{eq:7}
\ee
and
\be
M_{d}=\int_{0}^{1}dz \frac{z^{d/2-1}}{[1+z]^{d}}=
\frac{1}{2}\frac{\Gamma^{2}(d/2)}{\Gamma (d)} ~~~.
\label{eq:8}
\ee

The
exponent $\nu$ governing the algebraic component of the large
$x$ behavior is given  at ${\cal O}(2)$ by
\be
\nu =\omega^{2}2^{d+1}\left(
K_{d}+\frac{M_{d}}{3^{d/2+1}}\right) ~~~.
\ee
At lowest order one has the
OJK results $\lambda =\frac{d}{2}$ and $\nu =0$.  While $K_{d}$ and
$\omega$ can be worked out analytically for specific values of $d$,
the expressions are not very illuminating.  Numerical values for
$\lambda$, $\nu$ and $\omega$ are given in Tables I and II
along with other results for comparison.

\section{Review of Auxiliary field Methods}

How can one organize this problem in terms of a perturbation theory expansion?
Let us consider first the most direct method since the exercise
is
instructive and
suggests other approaches.
Let us ignore the fluctuation field $u$ in the decomposition of the
order-parameter field and replace the order-parameter field
$\psi$ with an auxiliary field
$m$ via the mapping
\be
\psi =\sigma [m]
\label{eq:mapping}
\ee
where $\sigma [m]$, as in TUG, satisfies the Euler-Lagrange equation for
the associated
stationary interface problem
\be
\frac{d^{2}\sigma }{dm^{2}}\equiv \sigma_{2}=V'[\sigma [m]] ~~~.
\label{eq:EL}
\ee
In this equation $m$ is taken to be the coordinate. It will generally
be useful to introduce the notation
\be
\sigma_{\ell}\equiv \frac{d^{\ell}\sigma}{dm^{\ell}} ~~~.
\ee
A key point in the introduction of the field $m$ is that the zeros
of $m$ locate the zeros of the order parameter and
give the positions of interfaces in the system.  Locally the magnitude
of $m$ gives the distance to the nearest defect.  As a system coarsens
and the distance between defects increases, the typical value of $m$
increases linearly with $L(t)$.

Inserting the mapping\cite{TUG3} given
by Eq.(\ref{eq:mapping})
into the equation of motion
for $\psi$ given by Eq.(\ref{eq:5}) one finds,
after using the chain-rule for differentiation, an equation for
$m (1)$:
\be
\sigma_{1}(1)\Lambda (1)m(1)=-\sigma_{2}(1)\left[1-(\nabla m(1))^{2}\right]
{}.
\ee
To get an idea of the nature of this equation take the special case of
a $\psi^{4}$ potential where one can solve for the mapping analytically
and obtain the usual interfacial kink form
\be
\sigma [m] = tanh (m/\sqrt{2})  ~~~.
\ee
It is easy to show for this particular potential, $
\sigma_{2}=-\sqrt{2}\sigma \sigma_{1}$ ,
and the equation of motion for $m$ can be written as
\be
\Lambda (1)m(1)
=\sqrt{2}\sigma (m(1))(1-(\nabla m(1))^{2}) ~~~.
\label{eq:BHeq}
\ee
The left-hand side looks like the
diffusion equation while the right-hand side has two types of nonlinearities.
The first nonlinearity is $\sigma (m)$ which looks like $sgn (m)$ in
the bulk.  The second
nonlinearity is given by the $(\nabla m)^{2}$ term.
Working along these lines,
Bray and Humayun\cite{BH} (BH) made the assumption
that one can  make a cleaver choice of the potential and replace
$-\sigma_{2} (m)/\sigma_{1} (m)$ by $m$ and then work with the equation
of motion
\be
\Lambda (1)m(1)
=m(1)(1-(\nabla m(1))^{2})  ~~~~.
\label{eq:BH}
\ee
This polynomial form looks promising .  If one makes the assumption
\be
(1-(\nabla m(1))^{2})\approx \frac{\kappa}{L^{2}(t_{1})},
\label{eq:BHapp}
\ee
then the resulting linearized equation
for $m$  is given by
\be
\Lambda (1)m(1)
=\frac{\kappa}{L^{2}(t_{1})}m(1) ~~~.
\label{eq:26}
\ee
This equation
is formally the same at the one we will find later in the
zeroth-order theory developed here and does corresponds to the
OJK theory.

It is not clear how to obtain systematic
corrections to Eq.(\ref{eq:26}).  Consider Eq.(\ref{eq:BH})
from the point of view of dimensional analysis.  Since
$m \approx L$, and $\Lambda \approx L^{-2}$, one has that the
left-hand side
Eq.(\ref{eq:BH}) is of ${\cal O}(L^{-1})$, while the right-hand
side is of
${\cal O}(L)$.  The only way that the right-hand side of
Eq.(\ref{eq:BH}) can be of ${\cal O}(L^{-1})$ is if there is the
additional constraint, given by  Eq.(\ref{eq:BHapp}), that is
enforced at all orders.
Formally this reduces to the statement that the
nonlinear interaction
\be
{\cal V}(1)=m(1)(1-(\nabla m)^{2})-\frac{\kappa}{L^{2}(t_{1})}m(1)
\ee
must lead to self-energy corrections which are of
${\cal O}(L^{-1})$.  Due to internal pairings of the
$\nabla m$ terms in such an expansion, this constraint can not be
enforced order by order and one arrives back at the dimensional
analysis argument discussed above.
Thus the assumption
given by Eq.(\ref{eq:BHapp}) is not self-consistent without
further development\cite{color}.

One ends up
concluding that the $m(\nabla m)^{2})$ nonlinearity is causing the
technical problems and the OJK theory is not the zeroth-order solution
of this problem {\it as~posed}.  Since there is reason to believe
that the OJK theory is
a good approximation to the original problem, then one
might conclude that the role of the $m(\nabla m)^{2})$
nonlinearity is technical and not crucial and one should reorganize
the calculation so that it plays a less prominent role.  Indeed
it is useful to
turn the argument around and suggest that it is the $\sigma [m]$
nonlinearity in Eq.(\ref{eq:BHeq})
 which is important in the equation of motion for $m$
and its role should be emphasized.

Let us back up a bit.  Suppose, instead of the rather rigid
mapping given by Eq.(\ref{eq:mapping}), we follow MVZ and
write
\be
\psi =\sigma [m] + u[m]
\label{eq:30b}
\ee
where $\sigma [m]$ is still the solution to the Euler-Lagrange
equation Eq.(\ref{eq:EL}) and $u[m]$ is to be determined.  Let us
substitute this
mapping into the equation of motion for the order parameter, use the
chain-rule as in leading to Eq.(\ref{eq:BHeq}), and obtain
\be
\Lambda (1)u(1)+\sigma_{1}(1)\Lambda (1)m(1)
\nonumber
\ee
\be
=-V'[\sigma (1)+u(1) ]
+\sigma_{2}(1)(\nabla m(1))^{2} ~~~.
\label{eq:30}
\ee
Notice that the perspective is different here.  This can be regarded as
an equation for the field $u$.  We then have the freedom to
assume that $m$ is driven by an equation of the type
given by Eq.(\ref{eq:BHeq}).  However we now choose this equation such that we
can find a self-consistent expansion about the OJK result.  To
begin we assume this equation of motion is of the form
\be
\Lambda (1)m(1)=\xi (t_{1}) \sigma (m(1))  ~~~.
\label{eq:30a}
\ee
Then this is just
Eq.(\ref{eq:BHeq}) with $(1-(\nabla m)^{2})$ replaced by the time dependent
quantity $\xi (t)$.
{}From dimensional analysis $m\approx L$, $\Lambda \approx L^{-2}$,
$\sigma\approx L^{ 0}$ so $\xi \approx L^{-1}$.
These restrictions are very important.
Using Eq.(\ref{eq:30a}) in Eq.(\ref{eq:30}) for $u$ leads to an
equation of motion for the field $u(1)$:
\be
\Lambda (1)u(1)=
-V'[\sigma (1)+u(1) ]+\sigma_{2}(1)(\nabla m(1))^{2}
\nonumber
\ee
\be
-\sigma_{1}(1) \xi (t_{1}) \sigma (m(1))  ~~~.
\ee
The most important aspect of the solution of the last equation
for $u=u[m]$ is that
\be
lim_{|m|\rightarrow \infty}u [m]=0 ~~~.
\ee
Indeed as far as the universal properties are concerned this is
almost all we need to proceed.  To understand that we can construct
a solution for u with this property let us consider the special case
where we have a $\psi ^{4}$ potential.  The equation for $u$ is given
then by
\be
\Lambda u+(3\sigma^{2}-1)u+3\sigma u^{2}+u^{3}
\nonumber
\ee
\be
=-\sigma_{2}\left[1-(\nabla m)^{2}\right]-\sigma_{1}
 \xi  \sigma  ~~~.
\label{eq:ueq}
\ee
In the limit of large $|m|$ the derivatives of $\sigma $ go
exponentially to zero and the right hand side of
Eq.(\ref{eq:ueq}) is
exponentially small.  Clearly we can construct a solution for
u where it is small and linearize the left hand side.  Remembering
that $\sigma ^{2} =1$ away from interfaces in the bulk we have
\be
\Lambda u+2u=
-\sigma_{2}(1-(\nabla m)^{2})-\sigma_{1}
 \xi ~ sgn (m)  ~~~.
\ee
Notice on the left-hand side that $u$ has acquired a
{\it mass}($=2$)
and in the long-time long-distance limit the term  where $u$ is
multiplied by a constant dominates the
derivative terms:
\be
2u=
-\sigma_{2}(1-(\nabla m)^{2}) ~~~.
\ee
We have dropped the term proportional to $\xi (t)$ since
it vanishes more rapidly than the
other terms at large times.  That the $u$ field picks up a mass in the
scaling limit can easily be seen to be a general feature of a wide
class of potentials where
$q_{0}^{2}=V''[\sigma =\pm \psi_{0}] > 0$ ~.
We have then on rather general principles that the field
$u$ must vanish rapidly as one moves into the bulk away from
interfaces.

One expects that the explicit construction of $u$ is rather involved and
depends on the details of the potential chosen.  If we restrict our
analysis to investigating
universal properties associated with bulk ordering
we will not need to know the statistics of $u$ explicitly.
If we are interested in determining interfacial properties then we
need to know $u$ in some detail.  Thus for example if we want to
determine the correlation function
\be
C_{\psi^{2}}(12)=<(\psi^{2}(1)-\psi_{0}^{2})(\psi^{2}(2)-\psi_{0}^{2})>
\ee
we will need to
know the statistics of the field $u$.  However
if we are interested in quantitites like
\be
C(12...n)=<\psi (1)\psi (2)\cdots \psi (n)>
\ee
where the points $12...n$ are not constrained to be close together,
then we do not need to know $u$ in detail.  Why is this?  Consider,
for example,
\be
C(12)=<\psi (1)\psi (2)>
=<\left[\sigma (1)+u(1)\right]\left[\sigma (2)+u(2)\right]> .
\nonumber
\ee
The point is that because the field $u(1)$ is nonzero only near
interfaces, the average
$<u(1)\sigma (2)>$ is down by a factor of $1/L^{2}$ relative to
$<\sigma (1)\sigma (2)>$.
Out in the bulk we can use dimensional analysis to make the following
estimates:
$(\nabla m)^{2}\approx {\cal O}(1)$~,
$\sigma_{2} \approx {\cal O}(L^{-2})$~,
$\sigma_{1}\xi (t) \sigma \approx {\cal O}(L^{-3})$~,
$\Lambda u\approx {\cal O}(L^{-4})$~,
where $\xi \approx L^{-1}$.
Using these results we find that the equation of motion for $u$
can be put into the form
\be
V'[\sigma +u ]=\sigma_{2}(\nabla m)^{2}
{}~~~~.
\ee
The usefulness of this equation is that it allows us to express
the equation of motion for the
order parameter, in the bulk, in terms of the field m,
\be
\Lambda \psi =-\sigma_{2}(\nabla m)^{2}
\label{eq:45}
{}~~~~.
\ee
We will demonstrate the usefulness of this result in section XI.
Our picture therefore has the $m$ field driving the order parameter
and the $m$ field satisfying the nonlinear Eq.(\ref{eq:30a}).

\section{Field Theory for Auxiliary Field}

Let us consider the field theory associated with the equation of motion
for $m(\vec{r},t)$.  Our  development will follow  the standard
Martin-Siggia-Rose\cite{MSR} method in its functional integral form
as developed by
DeDominicis and Peliti\cite{DP}.  The analysis begins with
the
basic equation of motion for the field $m$ given by Eq.(\ref{eq:30a}).
In the MSR method the field
theoretical development requires a doubling of operators to include the
field $M$ which is conjugate to $m$.
We also organize things so that the initial field $m_{0}(\vec{r})$ is
also treated as an independent field.
Thus it is assumed that $m$ is zero for $t < t_{0}$ and one must add
a term $\delta (t_{1}-t_{0})m_{0}(\vec{r}_{1})$ to the right-hand
side of Eq.(\ref{eq:30a}).

Following
standard procedures, averages of interest are given as functional
integrals over the
fields $m$, $M$ and $m_{0}$ weighted by an action
$A$:
\be
<m(1)m(2)\cdots  m(n) M(n+1)M(n+2)
\cdots M(n+\ell )>
\ee
\be
=\int {\cal D}m {\cal D}M{\cal D}m_{0}
m(1)m(2)\cdots  m(n)
\nonumber
\ee
\be
\times M(n+1)M(n+2)
\cdots M(n+\ell )
e^{A_{T}(m,M,m_{0})}/Z
\nonumber
\ee
where
\be
Z=\int {\cal D}m {\cal D}M{\cal D}m_{0}e^{A(m,M,m_{0})} ~~~.
\ee
The action takes the form
\be
A(m,M,m_{0})=-i\int d1M(1)
\nonumber
\ee
\be
\times\left[\Lambda (1)
m(1)-\xi(1)\sigma (1) -\delta (t_{1}-t_{0})m_{0}(1)\right]
\nonumber
\ee
\be
-\frac{1}{2}\int ~d^{d}r_{1}\int ~d^{d}r_{2}~m_{0}(\vec{r}_{1})
g^{-1}(\vec{r}_{1}-\vec{r}_{2})m_{0}(\vec{r}_{2})
\ee
where we use the notation,
$\int ~d1=\int ~dt_{1}d^{d}r_{1}$~,
and where we assume, as is appropriate in this case, that the initial field
is gaussian and has a variance given by
\be
<m_{0}(\vec{r}_{1})m_{0}(\vec{r}_{2})>=g(\vec{r}_{1}-\vec{r}_{2}) ~~~.
\ee
We will not have to be very specific about the form of the initial
correlation function $g$.  It will be very convenient to generate our
correlation functions as functional derivatives in terms of sources which
couple to the conjugate fields.  Thus we introduce
\be
S[h,H]
=exp\int d1 \left[h(1)m(1)+H(1)M(1)\right]
\ee
and define
\be
Z\left[h,H\right]=\int {\cal D}m {\cal D}M{\cal D}m_{0}
e^{A(m,M,m_{0})}S[h,H]
\nonumber
\ee
\be
\equiv
\int {\cal D}m {\cal D}M{\cal D}m_{0}
e^{A_{T}(m,M,m_{0})}
\ee
where the total action is defined
\be
A_{T}=A+\int d1 \left[h(1)m(1)+H(1)M(1)\right] ~~~.
\ee
The fundamental equations of motion are given by the identities
\be
\int {\cal D}m {\cal D}M{\cal D}m_{0}
\frac{\delta}{\delta M(1)}
e^{A_{T}(m,M,m_{0})}=0
\nonumber
\ee
\be
\int {\cal D}m {\cal D}M{\cal D}m_{0}
\frac{\delta}{\delta m(1)}
e^{A_{T}(m,M,m_{0})}=0
\nonumber
\ee
\be
\int {\cal D}m {\cal D}M{\cal D}m_{0}
\frac{\delta}{\delta m_{0}(1)}
e^{A_{T}(m,M,m_{0})}=0
\nonumber
\ee
which reduce to
\be
<\frac{\delta}{\delta M(1)}A_{T}(m,M,m_{0})>_{h}=0
\nonumber
\ee
\be
<\frac{\delta}{\delta m(1)}A_{T}(m,M,m_{0})>_{h}=0
\nonumber
\ee
\be
<\frac{\delta}{\delta m_{0}(1)}A_{T}(m,M,m_{0})>_{h}=0
\nonumber
\ee
where the subscript $h$ indicates that the average includes the
source fields $h$ and $H$:
\be
<...>_{h}=\frac{\int {\cal D}m {\cal D}M{\cal D}m_{0}e^{A_{T}(m,M,m_{0})}...}
{Z(h,H)}~~~.
\nonumber
\ee
Taking the functional derivative with respect to $M$ just
generates the original equation of motion  Eq.(\ref{eq:30a})
with an initial condition and a source term
\be
\frac{\delta}{\delta M(1)}A_{T}(m,M,m_{0})
\nonumber
\ee
\be
=-i\left[\Lambda (1)
m(1)-\xi(1)\sigma (1)
-\delta (t_{1}-t_{0})
m_{0}(1)\right]+H(1) ~~~.
\nonumber
\ee
The functional derivative with respect to $m$ is given by
\be
\frac{\delta}{\delta m(1)}A_{T}(m,M,m_{0})=
i\left[\tilde{\Lambda}(1)M(1)
+\xi(1)\sigma_{M}(1)
\right]+h(1)
\nonumber
\ee
where we have introduced the quantities
\be
\tilde{\Lambda}(1)=\frac{\partial}{\partial t_{1}}+\nabla^{2}_{1}
\nonumber
\ee
\be
\sigma_{M}(1)=\sigma_{1} (m(1))M(1)
\label{eq:67}
{}~~~.
\ee
Finally we need the derivative
\be
\frac{\delta}{\delta m_{0}(1)}A_{T}(m,M,m_{0})
\nonumber
\ee
\be
=iM(\vec{r}_{1},t_{0})
-\int d^{d}r_{2}g^{-1}(\vec{r}_{1}-\vec{r}_{2})
m_{0}(\vec{r}_{2})  ~~~.
\nonumber
\ee
Inserting the results of taking these derivatives into the averages
we obtain our fundamental
equations:
\be
-i\left[\tilde{\Lambda}(1)<M(1)>_{h}
+\xi(1)<\sigma_{M}(1)>_{h}
\right]=h(1)
\label{eq:73}
\ee
\be
i\left[\Lambda (1)
<m(1)>_{h}-\xi(1)<\sigma (1)>_{h}
-\delta (t_{1}-t_{0})
<m_{0}(1)>_{h}\right]
\nonumber
\ee
\be
=H(1) ~~~.
\label{eq:74}
\ee
and
\be
i<M(\vec{r}_{1},t_{0})>_{h}
=\int d^{d}r_{2}g^{-1}(\vec{r}_{1}-\vec{r}_{2})
<m_{0}(\vec{r}_{2})>_{h} ~~~.
\ee
The last equation allows one to solve for the average of the initial
field in terms of the MSR field $M$,
\be
<m_{0}(\vec{r}_{1})>_{h} =i\int d^{d}r_{2}g(\vec{r}_{1}-\vec{r}_{2})
<M(\vec{r}_{2},t_{0})>_{h} ~~~.
\ee
Eq.(\ref{eq:74}) can then be written in the form
\be
i\left[\Lambda (1)
<m(1)>_{h}-\xi(1)<\sigma (1)>_{h}\right]
\nonumber
\ee
\be
=-\int d2 ~\Pi_{0} (12)<M(2)>_{h}+H(1)
\label{eq:77}
\ee
where
\be
\Pi_{0} (12) \equiv \delta (t_{1}-t_{0})\delta (t_{1}-t_{2})
g(\vec{r}_{1}-\vec{r}_{2}) ~~~.
\ee
All correlation functions of interest can be
generated as functional derivatives of $<m(1)>_{h}$ or $<M(1)>_{h}$
with respect to $h(1)$ and $H(1)$.

In the limit in which  the source fields vanish, each term in the
two fundamental equations vanish.  Therefore it is derivatives of these
equations which are of interest.  Taking the functional derivative of
Eq.(\ref{eq:73})
with respect to $h(2)$ gives the equation for the response function
\be
G_{Mm}(12)=\frac{\delta}{\delta h(2)}<M(1)>_{h}
\ee
with
\be
-i\left[\tilde{\Lambda}(1)G_{Mm}(12)
+\xi(1)\frac{\delta <\sigma_{M} (1)>_{h}}{\delta h(2)}
\right]=\delta (12) ~~~.
\label{eq:80}
\ee
Taking the functional derivative of Eq.(\ref{eq:77})
with respect to $H(2)$ gives
\be
i\left[\Lambda (1)G_{mM}(12)
-\xi(1)\frac{\delta <\sigma (1)>_{h}}{\delta H(2)}
\right]
\nonumber
\ee
\be
=-\int d3 ~\Pi_{0} (13)G_{MM}(32)
+\delta (12) ~~~.
\label{eq:81}
\ee
Taking the functional derivative of Eq.(\ref{eq:77})
with respect to $h(2)$ gives
\be
i\left[\Lambda (1)G_{mm}(12)
-\xi(1)\frac{\delta <\sigma (1)>_{h}}{\delta h(2)}
\right]
\nonumber
\ee
\be
=-\int d3 ~\Pi_{0} (13)G_{Mm}(32) ~~~.
\ee
Note that Eqs.(\ref{eq:80}) and (\ref{eq:81}) are redundant
because of the relation
\be
G_{mM}(12)=G_{Mm}(21) ~~~.
\ee
Clearly we can go on and generate equations for all of the cumulants by
taking functional derivatives.  Let us introduce the notation that
$G_{A_{1},A_{2},....,A_{n}}(12...n)$ is the $n^{th}$ order cumulant
for the set of fields $\{A_{1},A_{2},....,A_{n}\}$ where field
$A_{1}$ has argument $(1)$, field $A_{2}$ has argument $(2)$, et cetera.
This notation is needed when we mix cumulants with $m$ and $M$.
As an example
\be
G_{Mmmm}(1234)=\frac{\delta^{3}<m(4)>_{h}}{\delta H(1)
\delta h(2)\delta h(3)} ~~~.
\ee
As a short hand for cumulants involving only $m$ fields we write
\be
G_{n}(12\cdots n)=\frac{\delta^{n-1}}{\delta h(n)\delta h(n-1)
\cdots \delta h(2)}<m(1)>_{h} ~~~.
\ee
The equations governing the $n^{th}$ order cumulants are given by
\be
-i\left[\tilde{\Lambda}(1)G_{Mm...m}(12...n)
+\hat{Q}_{n}(12...n)\right]=0
\label{eq:82}
\ee
and
\be
i\left[\Lambda (1)G_{n}(12...n)-Q_{n}(12...n)\right]
\nonumber
\ee
\be
=-\int
d\bar{1} ~\Pi_{0} (1\bar{1})G_{Mm...m}(\bar{1}2...n) ~~~.
\label{eq:83}
\ee
The $Q's$ are defined by
\be
\hat{Q}_{n}(12...n)=\xi(1)
\frac{\delta^{n-1}}{\delta h(n)\delta h(n-1)\cdots\delta h(2)}
<\sigma_{M}(1)>_{h}
\label{eq:88}
\ee
\be
Q_{n}(12...n)=\xi(1)
\frac{\delta^{n-1}}{\delta h(n)\delta h(n-1)\cdots\delta h(2)}
<\sigma (1)>_{h}.
\label{eq:89}
\ee
With this notation the equations determining the two-point functions can be
written as
\be
-i\left[\tilde{\Lambda}(1)G_{Mm}(12)
+\hat{Q}_{2}(12)\right]=\delta (12)
\label{eq:90}
\ee
\be
i\left[\Lambda (1)G_{2}(12)-Q_{2}(12)\right]
=-\int
d\bar{1} ~\Pi_{0} (1\bar{1})G_{Mm}(\bar{1}2)
\ee
We see that $G_{Mm...m}$ and
$G_{mm...m}$ are coupled.

The point now is to show that there is a consistent perturbation expansion
where the
higher-order cumulants are also of higher order in some ordering parameter.
To get started we need to express $\hat{Q}_{1}(12)$ and $Q_{1}(12)$
in terms of the fundamental cumulants of the $m$-field.  The first
step in this direction
is to show that these quantities can be expressed in terms of the
probability distribution
\be
P_{h}(x,1)=<\delta (x-m(1))>_{h}
=\frac{<\delta (x-m(1))S[h,H]>}{Z[h,H]}
\ee
where the averages without the subscript $h$ are weighted by $A$
not $A_{T}$ and the dependence on the source fields is explicit.
The average over $\sigma (m)$ can then be written
\be
<\sigma (m(1))>_{h}=\int dx \sigma (x)
P_{h}(x,1)
\ee
and
\be
Q_{1}(1)=\int dx ~
\xi (1)\sigma (x)~P_{h}(x,1) ~~~.
\ee
Evaluation of  $\hat{Q}_{1}(1)$ is only slightly more involved.  We have,
using Eq.(\ref{eq:67}), that
\be
<\sigma_{M} (1)>_{h}=<M(1)\sigma_{1}(m(1))>_{h}
\ee
\be
=\frac{1}{Z[h,H]}\frac{\delta}{\delta H(1)}
<\sigma_{1}(m(1))S[h,H]>
\nonumber
\ee
\be
=\frac{1}{Z[h,H]}\frac{\delta}{\delta H(1)}\left[
Z[h,H]\int dx ~\sigma_{1}(x)P_{h}(x,1)\right]
\nonumber
\ee
\be
=\left[<M(1)>_{h}+\frac{\delta}{\delta H(1)}\right]
\int dx ~\sigma_{1}(x)P_{h}(x,1)~~~.
\nonumber
\ee
Using this result in Eq.(\ref{eq:88}) with $n=1$,
\be
\hat{Q}_{1}(1)=\int ~dx ~ \xi (1)\sigma_{1} (x)
\left[<M(1)>_{h}
+\frac{\delta}{\delta H(1)}\right]
P_{h}(x,1).
\nonumber
\ee
Then any perturbation theory expansion for $P_{h}(x,1)$ will
lead immediately to an expansion for $\hat{Q}_{1}(1)$ and
$Q_{1}(1)$.  We can then obtain  $\hat{Q}_{n}$ and $Q_{n}$ by functional
differentiation.

\section{Perturbation Theory Expansion}

The perturbation theory expansion for $P_{h}(x,1)$ is straightforward.
Using the
integral representation for the $\delta$-function we have
\be
P_{h}(x,1)= \int\frac{dk}{2\pi}e^{-ikx}\Phi (k,h,1)
\label{eq:96}
\ee
where
\be
\Phi (k,h,1)=
<e^{{\cal H}(1)}>_{h}
\ee
and
${\cal H}(1)\equiv ikm(1)$~.
The average of the exponential is precisely of the form which can be
rewritten in terms of cumulants:
\be
\Phi (k,h,1)
=exp\left[\sum_{s=1}^{\infty}\frac{1}{s!}G_{{\cal H}}^{(s)}(1)\right]
\ee
where
$G_{{\cal H}}^{(s)}(1)$ is the $s^{th}$ order cumulant for the field
${\cal H}(1)$.
Since ${\cal H}(1)$ is proportional to $m(1)$ these are, up to factors
of $ik$ to the $s^{th}$ power, just the cumulants for the $m$
field:
\be
G_{{\cal H}}^{(s)}(1)=(ik)^{s}G_{s}(11\ldots 1) ~~~.
\ee
We can therefore write
\be
\Phi (k,h,1)
=exp\left[\sum_{s=1}^{\infty}\frac{(ik)^{s}}{s!}G_{s}(11...1)\right] ~~~.
\label{eq:103}
\ee

Consider
first the lowest-order contribution to $Q_{n}$ which does not vanish with the
external fields $h,H$:
\be
Q_{2}(12)=\int dx ~
\xi (1)\sigma (x)\frac{\delta}{\delta h(2)}P_{h}(x,1) ~~~.
\ee
We will assume, as we will show self-consistently, that  in zero external
field, $n^{th}$ order cumulants are
of order $\frac{n}{2}-1$ in an expansion parameter we will develop.
Expanding $\Phi (k,h,1)$ in powers of the cumulants with $n > 2$ and
keeping
terms up to the 4-point cumulant, we obtain
\be
P_{h}(x,1)=\left[1-\frac{1}{3!}G_{3}(111)\frac{d^{3}}{dx^{3}}
+\frac{1}{4!}G_{4}(1111)\frac{d^{4}}{dx^{4}}+\cdots\right]
\nonumber
\ee
\be
\times P_{h}^{(0)}(x,1)
\ee
where
\be
P_{h}^{(0)}(x,1)=\int\frac{dk}{2\pi} \Phi_{0}(k,h,1)e^{-ikx}
\ee
and
\be
\Phi_{0}(k,h,1)=e^{ikG_{1}(1)}
e^{-\frac{1}{2}k^{2}G_{2}(11)} ~~~.
\ee
Then,
after taking the derivative with
respect to $h(2)$, setting the external fields to zero, and
neglecting all cumulants with $n > 2$, we obtain
\be
\Phi_{0}(k,h=0,1)=e^{-\frac{1}{2}k^{2}S_{2}(1)} ~~~,
\ee
and
\be
Q_{2}^{(0)}(12)=\int dx ~
\xi (1)\sigma (x)
\int\frac{dk}{2\pi}e^{-ikx}ik G_{2}(12)e^{-\frac{1}{2}k^{2}S_{2}(1)}
\nonumber
\ee
where we have defined in zero external field
\be
S_{2}(1)\equiv G_{2}(11)=<m^{2}(1)> ~~~.
\ee
There are several points to make here.  Let us begin with the
separation of the factor $\xi (1)\sigma (x)$ into a piece
which contributes to the bulk universal properties and a
part which does not.  Note that in general we can write $\sigma (x)$
in the form
\be
\sigma (x) =\psi_{0}~ sgn(x) +\tilde{\sigma} (x)
\ee
where $\psi_{0}$ is the bulk ordered value of the magnitude of the
order parameter and $\tilde{\sigma} (x)$ goes to zero exponentially
as $|x|\rightarrow \infty$.  For the $\psi^{4}$ potential
$\psi_{0}=1$ and
\be
\tilde{\sigma} (x)=-sgn(x)\left[\frac{2}{e^{\sqrt{2}|x|}+1}\right]~~~.
\ee
This means that $Q_{2}^{(0)}(12)$ can be written as the sum of two
pieces:
\be
Q_{2}^{(0)}(12)=Q_{2}^{(0,B)}(12)+Q_{2}^{(0,N)}(12)
\ee
where
\be
Q_{2}^{(0,B)}(12)=\xi (1)\psi_{0}G_{2}(12)
\ee
\be
\times\int dx
{}~ sgn (x)
\int\frac{dk}{2\pi}ik e^{-ikx} e^{-\frac{1}{2}k^{2}S_{2}(1)}
\ee
and
\be
Q_{2}^{(0,N)}(12)=G_{2}(12)\int dx ~
\xi (1)\tilde{\sigma} (x)~
\int\frac{dk}{2\pi}ik e^{-ikx} e^{-\frac{1}{2}k^{2}S_{2}(1)} .
\nonumber
\ee
In both contributions we have the integral
\be
\int\frac{dk}{2\pi}e^{-ikx}ik e^{-\frac{1}{2}k^{2}S_{2}(1)}
=-\frac{d}{dx}\int\frac{dk}{2\pi} e^{-ikx} e^{-\frac{1}{2}k^{2}S_{2}(1)}
\nonumber
\ee
\be
=-\frac{d}{dx}\Phi_{0}(x,1)
\ee
where $\Phi_{0}(x,1)$ is the Fourier transform of $\Phi_{0}(k,h=0,1)$:
\be
\Phi_{0}(x,1)=\int\frac{dk}{2\pi}e^{-ikx} e^{-\frac{1}{2}k^{2}S_{2}(1)}
=\frac{e^{-\frac{x^{2}}{2S_{2}(1)}}}{\sqrt{2\pi S_{2}(1)}} ~~~.
\ee
In evaluating $Q_{2}^{(0,N)}(12)$ we can take the derivative of
$\Phi_{0}(x,1)$ and expand in inverse powers of $S_{2}(1)$ to obtain
to leading order
\be
Q_{2}^{(0,N)}(12)=G_{2}(12)\int dx
\xi (1)\tilde{\sigma} (x)
\frac{x}{\sqrt{2\pi}S_{2}^{3/2}(1)}
\nonumber
\ee
where it is crucial that $\tilde{\sigma} (x)$
vanish for large $|x|$ so that the $x$-integral exists.  Turning to
$Q_{2}^{(0,B)}(12)$ we have
\be
Q_{2}^{(0,B)}(12)=\xi (1)\psi_{0}G_{2}(12)\int dx
{}~sgn (x)\left[-\frac{d}{dx}\Phi_{0}(x,1)\right] .
\nonumber
\ee
In this case we integrate by parts in the integral over $x$ and use
$\frac{d}{dx} sgn (x) =2\delta (x)$
to obtain
\be
Q_{2}^{(0,B)}(12)=\xi (1)\psi_{0}G_{2}(12)
2\Phi_{0}(0,1)
\nonumber
\ee
\be
=\xi (1)\psi_{0}G_{2}(12)\sqrt{\frac{2}{\pi S_{2}(1)}}~~~.
\ee
The key observation
is that $Q_{2}^{(0,N)}(12)$ is down by a factor of $1/L^{2}$ relative to
$Q_{2}^{(0,B)}(12)$.  It should be clear that this is a general result
which will hold order by order in perturbation theory.  As far as the
bulk ordering properties are concerned we can replace
$\sigma (x)\rightarrow \psi_{0}~ sgn(x)$ ~.
Thus we could have started with the equation of motion for the field
$m$
\be
\Lambda (1) m(1)=\xi (1)\psi_{0}~ sgn (m(1))
\ee
if we focus only on bulk ordering properties\cite{uniarg}.

Turning to the other nonlinear quantity $\hat{Q}_{2}$, we have in
general
\be
\hat{Q}_{2}(12)=\int ~dx ~\xi (1)\sigma_{1} (x)
\label{eq:135}
\ee
\be
\times\left[G_{Mm}(12)+<M(1)>_{h}
\frac{\delta}{\delta h(2)}
+\frac{\delta^{2}}{\delta h(2)\delta H(1)}\right]
P_{h}(x,1) .
\nonumber
\ee
Clearly term by term in our expansion we will find that the leading
contribution comes from
$\sigma_{1} (x)
\rightarrow \psi_{0}2\delta (x)$
with the remaining terms  leading to contributions which are of
higher order in $1/L$.  We therefore need only consider
\be
\hat{Q}_{2}(12)=\xi (1)\psi_{0}\int ~dx  2\delta (x)
\nonumber
\ee
\be
\times\left[G_{Mm}(12)+<M(1)>_{h}
\frac{\delta}{\delta h(2)}
+\frac{\delta^{2}}{\delta h(2)\delta H(1)}\right]
P_{h}(x,1)
\nonumber
\ee
if we are only interested in the bulk universal properties.

A key observation is that as we analyze contributions to
$Q_{n}$ or $\hat{Q}_{n}$ we will find that each term consists of
products of correlation functions and response functions with
legs tied together by factors defined by
\be
\phi_{p}(1)\equiv\int dx~sgn (x) \int \frac{dk}{2\pi}
ik^{2p+1}e^{-ikx}\Phi_{0}(k,1)
\ee
\be
=2\int \frac{dk}{2\pi}k^{2p}e^{-\frac{1}{2}k^{2}S_{2}(1)}
=\left(-2\frac{d}{dS_{2}(1)}\right)^{p}\phi_{0}(1)
\nonumber
\ee
where we have used an integration by parts in going from the
first to the second line and defined
\be
\phi_{0}(1)=2\int \frac{dk}{2\pi}e^{-\frac{1}{2}k^{2}S_{2}(1)}
=\sqrt{\frac{2}{\pi S_{2}(1)}}~~~.
\ee
Each term in the perturbation theory expansion for
$Q_{n}$ or $\hat{Q}_{n}$
will be
proportional to factors of $\phi_{p}$.
The perturbation expansion is ordered by the
sum of the labels $p$ on $\phi_{p}$.  Thus a contribution with insertions
$\phi_{1}\phi_{2}\phi_{1}$ ~,
each factor
typically associated with different times,
is of ${\cal O}(4)$.
We shall refer to this expansion as the {\it $\phi$-expansion}.
It should be emphasized that at this stage that this is a {\it formal}
expansion.  At order $n$ it is true that
$\phi_{p}\approx L^{-(2p+1)}$
which is small, however it will be multiplied, depending on the quantity
expanded, by positive factors of $L(t)$ such that each term in the
expansion in $\phi_{p}$ has the same
overall leading power with respect to $L(t)$.

To see how this expansion works let us consider first the two-point
quantity $Q_{2}(12)$, defined by
\be
Q_{2}(12)=\int~dx~\xi (1) sgn(x)
\frac{\delta P_{h}(x,1)}{\delta h(2)} ~~~.
\ee
Using Eqs.(\ref{eq:96}) and (\ref{eq:103}) and taking the derivatives
with respect to $h(2)$,
we find, in the case of zero external
fields,
\be
Q_{2}(12)=\xi(1)\int~dx ~sgn(x)\int~\frac{dk}{2\pi}e^{-ikx}\Phi (k,h=0,1)
\nonumber
\ee
\be
\times\sum_{s=0}^{\infty}\frac{(ik)^{2s+1}}{(2s+1)!}
G_{2s+2}(11...12) ~~~.
\ee
Since all odd cumulants vanish in
the case of zero external fields  and
\be
\Phi (k,h=0,1)=exp\left[\sum_{s=1}^{\infty}\frac{(-1)^{s}k^{2s}}{(2s)!}
S_{2s}(1)\right]
\ee
where
\be
S_{2s}(1)=G_{2s}(11...1) ~~~.
\ee
Let us define the set of vertices
\be
{\cal V}_{p}(1)=
\int dx~sgn (x) \int \frac{dk}{2\pi}
ik^{2p+1}e^{-ikx}\Phi(k,h=0,1) ~~~,
\label{eq:141}
\ee
which reduces, after following the same set of steps in reducing
the original expression for $\phi_{p}$, to
\be
{\cal V}_{p}(1)=2 \int \frac{dk}{2\pi} k^{2p}\Phi(k,h=0,1)
\ee
which is independent of position.  Then the  quantity $Q_{2}(12)$,
which appears in the
equation of motion for $G_{2}(12)$, is given in the form
\be
Q_{2}(12)=\xi(1)\psi_{0} \sum_{s=0}^{\infty}\frac{(-1)^{s}}{(2s+1)!}
G_{2s+2}(11...12)
\nonumber
\ee
\be
\times\int~dx ~sgn(x)\int~\frac{dk}{2\pi}e^{-ikx}
ik^{2s+1}\Phi (k,1)
\nonumber
\ee
\be
=\xi(1) \sum_{s=0}^{\infty}\frac{(-1)^{s}}{(2s+1)!}
{\cal V}_{s}(1)
G_{2s+2}(11...12)
\label{eq:131}
\ee
where we have used the definition of ${\cal V}_{s}(1)$ given by
Eq.(\ref{eq:141})
in the last step.

It should be clear that the vertices ${\cal V}_{s}(1)$ are of at least
${\cal O}(s)$ in the $\phi$-expansion.  By direct expansion of
$\Phi (k,h=0,1)$ about $\Phi_{0} (k,h=0,1)$ we obtain
\be
{\cal V}_{s}(1)=\phi_{s}(1)+\frac{S_{4}(1)}{4!}\phi_{s+2}(1)
-\frac{S_{6}(1)}{6!}\phi_{s+3}(1)
\nonumber
\ee
\be
+\phi_{s+4}(1)\left[\frac{S_{4}^{2}(1)}{2(4!)^{2}}+\frac{S_{8}(1)}{8!}\right]
+\cdots ~~~.
\ee
Then since we will find $S_{\ell}(1)\approx {\cal O}(\frac{\ell}{2}-1)$,
the terms in the expansion for ${\cal V}_{s}$ are of ${\cal O}(s)$,
${\cal O}(s+3)$, ${\cal O}(s+5)$ and ${\cal O}(s+6)$ respectively.

Let us turn next to $\hat{Q}_{2}(12)$ given by Eq.(\ref{eq:135}).
In the limit
of zero external fields we can set the term proportional to
$<M(1)>$ to zero and  then evaluate the second derivatives,
$\delta^{2} P_{h}(x,1)/\delta h(2)\delta H(1)$.  After a
significant amount of algebra we obtain\cite{causality}:
\be
\hat{Q}_{2}(12)
=\int ~dx 2\xi (1)\psi_{0}\delta (x)
\int~\frac{dk}{2\pi}e^{-ikx}\Phi (k,1)
\nonumber
\ee
\be
\times\left[ G_{Mm}(12)
+\sum_{s=1}^{\infty}\frac{(ik)^{s}}{s!}
G^{(s+2)}_{mm...mMm}(11...112 )\right]
\nonumber
\ee
\be
=\int ~dx 2\xi (1)\psi_{0}\delta (x)
\int~\frac{dk}{2\pi}e^{-ikx}\Phi (k,1)
\nonumber
\ee
\be
\times\sum_{s=0}^{\infty}\frac{(ik)^{s}}{s!}
G^{(s+2)}_{mm...mMm}(11...112 )
\nonumber
\ee
\be
=\xi (1)\psi_{0}\sum_{s=0}^{\infty}\frac{(-1)^{s}}{(2s)!} {\cal V}_{s}(1)
G^{(2s+2)}_{mm...mMm}(11...112 )  ~~~.
\label{eq:133}
\ee

Before going on to discuss the structure of the perturbation theory
at higher order let us made sure the theory is sensible at zeroth order
where, from Eqs.(\ref{eq:131}) and (\ref{eq:133})
\be
Q_{2}^{(0)}(12)=\xi(1)\psi_{0}\phi_{0}(1)G_{2}(12)
\label{eq:154}
\ee
\be
\hat{Q}_{2}^{(0)}(12)=\xi(1)\psi_{0}\phi_{0}(1)G_{Mm}(12) ~~~.
\label{eq:153}
\ee

\section{Zeroth Order Theory For Two Point Correlation Functions}

The equations of motion at zeroth order for the two-point correlation
function is given by Eqs.(\ref{eq:82}) and (\ref{eq:83})
with $\hat{Q}_{2}$ and
$Q_{2}$ replaced by Eqs.(\ref{eq:153}) and (\ref{eq:154}).  Thus we have
\be
-i\left[\tilde{\Lambda}(1)+\omega_{0}(1)\right]G_{Mm}^{(0)}(12)
=\delta (12)
\ee
\be
i\left[\Lambda (1)-\omega_{0}(1)\right]G_{2}^{(0)}(12)
=-\int
d\bar{1} ~\Pi_{0} (1\bar{1})G_{Mm}^{(0)}(\bar{1}2)
\ee
where we have defined
\be
\omega_{0} (1) =\xi(1)\psi_{0}\phi_{0}(1) ~~~.
\ee

The first step in the solution to these equations is to  Fourier transform
over space.  Taking the equation for the response function first,
we obtain
\be
-i\left[\frac{\partial}{\partial t_{1}}-q^{2}+\omega_{0}(t_{1})\right]
G_{Mm}^{(0)}(q,t_{1}t_{2})
=\delta (t_{1}-t_{2}) ~~~.
\ee
This first-order differential equation has the solution
\be
G_{Mm}^{(0)}(q,t_{1}t_{2})=-i\theta (t_{2}-t_{1})
exp\left[\int_{t_{2}}^{t_{1}}~d\tau\left(q^{2}-\omega_{0} (\tau )\right)\right]
\ee
\be
=-i\theta (t_{2}-t_{1})R(t_{2},t_{1})e^{-q^{2}(t_{2}-t_{1})}
\nonumber
\ee
and we have defined
\be
R(t_{1},t_{2})=e^{\int_{t_{2}}^{t_{1}}~d\tau\omega_{0} (\tau )}  ~~~.
\label{eq:161}
\ee
Taking the inverse Fourier transform, using
\be
\int ~\frac{d^{d}q}{(2\pi )^{d}}e^{i\vec{q}\cdot\vec{r}}
e^{-q^{2}(t_{1}-t_{2})}
=\frac{e^{-\frac{r^{2}}{4(t_{1}-t_{2})}}}{[4\pi (t_{1}-t_{2})]^{d/2}}
{}~~~,
\ee
we obtain
\be
G_{Mm}^{(0)}(r,t_{1}t_{2})=-i\theta (t_{2}-t_{1})R(t_{2},t_{1})
\frac{e^{-\frac{r^{2}}{4(t_{2}-t_{1})}}}{[4\pi (t_{2}-t_{1})]^{d/2}}
{}~~~.
\ee
It is straightforward to show explicitly the result we expect from
symmetry considerations:
\be
G_{mM}^{(0)}(r,t_{1}t_{2})=-i\theta (t_{1}-t_{2})R(t_{1},t_{2})
\frac{e^{-\frac{r^{2}}{4(t_{1}-t_{2})}}}{[4\pi (t_{1}-t_{2})]^{d/2}}
{}~~~.
\ee
Let us turn our attention to the correlation function.
It is useful to introduce the
inverse propagators
\be
G_{Mm}^{0,-1}(12)=-i\left[\tilde{\Lambda} (1)+\omega_{0}(1)\right]\delta (12)
\ee
\be
G_{mM}^{0,-1}(12)=i\left[\Lambda (1)-\omega_{0}(1)\right]\delta (12)
\ee
which  allows one to write the equation for the correlation function in
the form
\be
\int~d\bar{1}G_{mM}^{0,-1}(1\bar{1})G_{2}^{(0)}(\bar{1}2)
=-\int
d\bar{1} ~\Pi_{0} (1\bar{1})G_{Mm}^{(0)}(\bar{1}2)
{}.
\ee
Multiplying from the left by $G_{mM}^{(0)}$ and using the definition
of the inverse gives the symmetric form:
\be
G_{2}^{(0)}(12) =-\int~d\bar{1}\int~d\bar{2}
{}~G_{mM}^{(0)}(1\bar{1})G_{mM}^{(0)}(2\bar{2})\Pi_{0} (\bar{1}\bar{2})
{}.
\label{eq:164}
\ee
Taking the Fourier transform and
inserting the results for the propagators and $\Pi_{0}$ we obtain
\be
G_{2}^{(0)}(q,t_{1}t_{2})=\theta (t_{1}-t_{0})\theta (t_{2}-t_{0})
e^{-q^{2}(t_{1}+t_{2}-2t_{0})}
\nonumber
\ee
\be
\times R(t_{1},t_{0})R(t_{2},t_{0})\tilde{g}(q)
\label{eq:174}
\ee
where $\tilde{g}(q)$ is the Fourier transform of the initial
correlation function.
Henceforth we will suppress writing the step functions in the correlation
function.  Inverting the Fourier transform we obtain
\be
G_{2}^{(0)}(r,t_{1}t_{2})=R(t_{1},t_{0})R(t_{2},t_{0})
\int \frac{d^{d}q}{(2\pi )^{d}}e^{i\vec{q}\cdot\vec{r}}\tilde{g}(q)
e^{-2q^{2}T}
\label{eq:170}
\ee
where it is convenient to introduce
\be
T=\frac{t_{1}+t_{2}}{2}-t_{0} ~~~.
\ee

While we are primarily interested in the long-time scaling
properties of our system, we can retain some control over the
influence of initial conditions and still be able to carry out the
analysis analytically if we introduce the initial condition
\be
\tilde{g}(q)=g_{0}e^{-\frac{1}{2}(q\ell )^{2}}
\ee
or
\be
g(\vec{r})=g_{0}
\frac{e^{-\frac{1}{2}(r/\ell )^{2}}}{(2\pi \ell^{2})^{d/2}} ~~~.
\ee
Inserting this form into Eq.(\ref{eq:170}) and doing the wavenumber
integration we obtain
\be
G_{2}^{(0)}(\vec{r},t_{1}t_{2})
\label{eq:175a}
\ee
\be
=R(t_{1},t_{0})R(t_{2},t_{0})
\frac{g_{0}}{[2\pi (\ell^{2}+4T)]^{d/2}}
e^{-\frac{1}{2}r^{2}/(\ell^{2}+4T)} .
\nonumber
\ee
In the long-time limit this reduces to
\be
G_{2}^{(0)}(r,t_{1}t_{2})=R(t_{1},t_{0})R(t_{2},t_{0})g_{0}
\frac{e^{-r^{2}/8T}}{(8\pi T)^{d/2}}  ~~~.
\label{eq:180}
\ee

Let us turn now to the quantity $R(t_{1},t_{2})$ defined by
Eq.(\ref{eq:161}).   We have assumed that $\xi (t)\approx 1/L(t)$
and we have $\phi_{0}\approx 1/L(t)$, so we can write for long
times
\be
\omega_{0} (t)=\xi (t)\psi_{0}\phi_{0}(t)=\frac{\omega}{t_{c}+t}
\label{eq:176}
\ee
where $\omega$ is a constant we will determine and $t_{c}$ is a
short-time cut-off which depends on details of the earlier-time
evolution.  Evaluating the integral
\be
\int_{t_{2}}^{t_{1}}~d\tau \omega_{0}(\tau )=
\int_{t_{2}}^{t_{1}}~d\tau\frac{\omega}{t_{c}+\tau }
=\omega~ln\left(\frac{t_{1}+t_{c}}{t_{2}+t_{c}}\right) ~~~,
\ee
we obtain
\be
R(t_{1},t_{2})=\left(\frac{t_{1}+t_{c}}{t_{2}+t_{c}}\right)^{\omega}~~~.
\ee
Inserting this result back into Eq.(\ref{eq:180}) leads to the
expression
for the correlation function
\be
G_{2}^{(0)}(r,t_{1}t_{2})=g(0)
\left(\frac{t_{1}+t_{c}}{t_{0}+t_{c}}\right)^{\omega}
\left(\frac{t_{2}+t_{c}}{t_{0}+t_{c}}\right)^{\omega}
\frac{e^{-r^{2}/8T}}{(8\pi T)^{d/2}}~~~.
\ee
If we are to have a self-consistent scaling equation then the
autocorrelation function $(r=0)$, at large equal times
$t_{1}=t_{2}=t$, given by
\be
S_{2}^{(0)}(t)=g_{0}\left(\frac{t}{t_{0}+t_{c}}\right)^{2\omega}
\frac{1}{(8\pi t)^{d/2}}
\nonumber
\ee
\be
=t^{2\omega-d/2}\frac{1}{(t_{0}+t_{c})^{2\omega}}
\frac{g_{0}}{(8\pi )^{d/2}} ~~~,
\ee
must have the form $S_{2}^{(0)}(t)=A_{0}t$.  Comparing we see the
exponent $\omega$ must be given by
\be
\omega=\frac{1}{2}(1+\frac{d}{2})
\label{eq:161a}
\ee
and the amplitude by
\be
A_{0}=\frac{1}{(t_{0}+t_{c})^{2\omega}}
\frac{g_{0}}{(8\pi )^{d/2}}
{}~~~.
\label{eq:184}
\ee

One question which arises is whether the time dependence in
$\omega_{0}(t)$ should be viewed as externally driven or
self-consistently
developed.  Rephrasing this question:  Does the time
evolution for $m(t)$ evolve out of the early-time instability?
It is
instructive to see that this follows in a natural fashion.  Suppose
instead of Eq.(\ref{eq:176}) we assume
\be
\omega_{0}(t)=\frac{\bar{\omega}}{S_{2}(t)}
\label{eq:185}
\ee
which is consistent with
\be
\psi_{0}\xi (t)=\frac{\omega_{0}(t)}{\phi_{0}(t)}
=\bar{\omega}\sqrt{\frac{\pi}{2S_{2}(t)}}
\ee
since $\phi_{0}(t)=\sqrt{2/\pi S_{2}(t)}$.  Inserting this result
for $\omega_{0}(t)$ into Eq.(\ref{eq:161}) and
Eq.(\ref{eq:170}) with
$r=0$, one finds at equal times $t_{1}=t_{2}=t$, a self-consistent
equation for the zeroth-order quantity $S_{2}^{(0)}(t)$:
\be
S_{2}^{(0)}(t)=exp\left(2\bar{\omega}\int_{t_{0}}^{t}
\frac{d\tau}{S_{2}^{(0)}(\tau)}\right)
\frac{S_{2}^{(0)}(t_{0})}{\left[1+\alpha (t-t_{0})\right]^{d/2}}
\label{eq:187}
\ee
where
\be
S_{2}^{(0)}(t_{0})=\frac{g_{0}}{(2\pi \ell^{2})^{d/2}}
\ee
and
\be
\alpha\equiv 4/\ell^{2}~~~.
\ee
This equation is solved in the Appendix  with the result
\be
S_{2}^{(0)}(t)=\left(S_{2}^{(0)}(t_{0})
-\frac{2\bar{\omega}}{\alpha (1+d/2)}\right)[1+\alpha (t-t_{0})]^{-d/2}
\nonumber
\ee
\be
+\frac{2\bar{\omega}}{\alpha (1+d/2)}[1+\alpha (t-t_{0})]
\ee
and $\bar{\omega}$ is still undetermined.  There are several
interesting results which follow.
If we look at the long-time limit of $\omega_{0}(t)$ we obtain
\be
\omega_{0}^{-1}(t)=\frac{S_{2}^{(0)}(t)}{\bar{\omega}}=
\frac{2}{\alpha (1+d/2)}(1+\alpha (t-t_{0})~~~.
\ee
If we compare with Eq.(\ref{eq:176}), we regain Eq.(\ref{eq:161a}) and
find
\be
t_{c}=\frac{1}{\alpha}-t_{0} ~~~.
\label{eq:198}
\ee
Thus our procedure is consistent and the system governed by
Eqs.(\ref{eq:174}) and Eqs.(\ref{eq:185}) is driven to grow as desired.
Since $S_{2}^{(0)}(t)$ is now a known function of time we can
use Eq.(\ref{eq:187}) to evaluate $R(t,t_{0})$ and obtain the complete
solution to this zeroth-order problem over the entire time
regime
\be
G_{2}^{(0)}(\vec{r},t_{1}t_{2})=
\frac{g_{0}}{[2\pi (\ell^{2}+4T)]^{d/2}}
e^{-\frac{1}{2}r^{2}/(\ell^{2}+4T)}
\nonumber
\ee
\be
\times
\left[\frac{S_{2}^{(0)}(t_{1})}{S_{2}^{(0)}(t_{0})}
\frac{S_{2}^{(0)}(t_{2})}{S_{2}^{(0)}(t_{0})}
(1+\alpha (t_{1}-t_{0}))^{d/2}(1+\alpha (t_{2}-t_{0}))^{d/2}\right]^{1/2}
{}.
\nonumber
\ee
In the long-time limit this can be written, using the definition
Eq.(\ref{eq:198}) for $t_{c}$, in the form
\be
G_{2}^{(0)}(r,t_{1}t_{2})=A_{0}
\left[\left(t_{1}+t_{c}\right)
\left(t_{2}+t_{c}\right)\right]^{\frac{1}{2}(1+\frac{d}{2})}
\frac{e^{-r^{2}/8T}}{ T^{d/2}}
\ee
with
\be
A_{0}=\frac{2\bar{\omega}}{1+d/2} ~~~.
\ee
Notice that this result for $A_{0}$ differs from the result given by
Eq.(\ref{eq:184}) and does not depend on the initial conditions.
This is evidence that the coefficient $A_{0}$ is nonuniversal.
Note, if we choose to  enforce the condition $<(\nabla m)^{2}>=1$
at late times (see Eq.(\ref{eq:BHapp}))
we can fix the parameter $\bar{\omega}$ at this order
with
the result
\be
\bar{\omega}= \frac{2}{d}(1+\frac{d}{2}) ~~~.
\ee

The general expression for the correlation function
can be rewritten in the convenient form
\be
G_{2}^{(0)}(r,t_{1}t_{2})=\sqrt{S_{2}^{(0)}(t_{1})S_{2}^{(0)}(t_{2})}
\Phi_{0}(t_{1}t_{2})
e^{-\frac{1}{2}r^{2}/(\ell^{2}+4T)}
\ee
where, using Eq.(\ref{eq:198}) for $1/\alpha $,
\be
\Phi_{0}(t_{1}t_{2})=\left(\frac{\sqrt{(t_{1}+t_{c})
(t_{2}+t_{c})}}
{ T+t_{c}+t_{0}}
\right)^{d/2}~~~.
\ee
The nonequilibrium exponent is defined in the long-time limit by
\be
\frac{G_{2}^{(0)}(0,t_{1}t_{2})}
{\sqrt{S_{2}^{(0)}(t_{1})S_{2}^{(0)}(t_{2})}}=
\left(\frac{\sqrt{(t_{1}+t_{c})(t_{2}+t_{c})}}
{T+t_{0}+t_{c}}\right)^{\lambda_{m}}
\label{eq:178}
\ee
and we obtain the OJK result
\be
\lambda_{m} =\frac{d}{2}~~~.
\ee
We put the subscript $m$ on $\lambda$ to indicate the exponent associated
with $G_{2}(12)$.  In general $G_{2}(12)$ and the order parameter
correlation function  $C(12)$ need not
share the same exponents.

Looking at equal times we have that
\be
f_{0}(x)=\frac{G_{2}^{(0)}(r,tt)}{S_{2}^{(0)}(t)}=e^{-x^{2}/2}
\ee
where the scaled length is defined by
$\vec{x}=\vec{r}/4t$~.
$f_{0}(x)$
is just the well known OJK result for the scaled auxiliary correlation
function.

\section{ Perturbation Theory for Four-Point Cumulants}

In order to see how things go at higher order in perturbation theory it
is useful to look at the lowest nonzero approximation
for the four-point cumulants which enters
into the ${\cal O}(2)$ correction for the two-point cumulant.
In order to compute $G_{4}(1234)$ and $G_{Mmmm}(1234)$ to ${\cal O}(1)$
in perturbation theory, we see from Eqs.(\ref{eq:82}) and (\ref{eq:83})
that we must evaluate
\be
Q_{4}(1234)=
\frac{\delta^{3}}{\delta h(4)\delta h(3)\delta h(2)}
\xi(1)\psi_{0}\int ~dx ~sgn(x)P_{h}(x,1),
\nonumber
\ee
and
\be
\hat{Q}_{4}(1234)=
\frac{\delta^{2}}{\delta h(4)\delta h(3)}\hat{Q}_{2}(12)
\nonumber
\ee
\be
=\frac{\delta^{2}}{\delta h(4)\delta h(3)}
\int ~dx 2\xi (1)\delta (x)
\nonumber
\ee
\be
\times\left[ G_{Mm}(12)
+<M(1)>_{h}\frac{\delta}{\delta h(2)}
+\frac{\delta^{2}}{\delta h(2)\delta H(1)}\right]
P_{h}(x,1)
\nonumber
\ee
\be
=\int ~dx 2\xi (1)\delta (x)
\Bigg[G_{Mmmm}(1234)P_{h}(x,1)
\nonumber
\ee
\be
+ G_{Mm}(12)\frac{\delta^{2}}{\delta h(4)\delta h(3)}P_{h}(x,1)
\nonumber
\ee
\be
+ G_{Mm}(13)\frac{\delta^{2}}{\delta h(2)\delta h(4)}P_{h}(x,1)
\nonumber
\ee
\be
+ G_{Mm}(14)\frac{\delta^{2}}{\delta h(2)\delta h(3)}P_{h}(x,1)
\nonumber
\ee
\be
+\frac{\delta^{4}}{\delta h(4)\delta h(3)\delta h(2)\delta H(1)}
P_{h}(x,1)\Bigg]
\ee
to ${\cal O}(1)$.
Assuming, as we will show
self-consistently, that $G_{n}\approx{\cal O}(\frac{n}{2}-1)$
and ${\cal V}_{n}\approx{\cal O}(n)$,  one easily finds that
the ${\cal O}(1)$ contributions to $Q_{4}$ and $\hat{Q}_{4}$
are given by
\be
Q_{4}^{(1)}(1234)=\omega_{0}(1)G_{4}(1234)
\nonumber
\ee
\be
-\omega_{1}(1) G_{2}(14)G_{2}(13)G_{2}(12)
\label{eq:203}
\ee
and
\be
\hat{Q}_{4}^{(1)}(1234)=\omega_{0}(1)G_{Mmmm}(1234)
\label{eq:204}
\ee
\be
-\omega_{1}(1)\Bigg[G_{Mm}(12)
G_{2}(13)G_{2}(14)
+G_{Mm}(13)
G_{2}(12)G_{2}(14)
\nonumber
\ee
\be
+G_{Mm}(14)
G_{2}(12)G_{2}(13)+G_{2}(14)G_{Mmmm}(1123)
\nonumber
\ee
\be
+G_{2}(13)G_{Mmmm}(1124)
+G_{2}(12)G_{Mmmm}(1134)\Bigg]
\nonumber
\ee
where we have introduced the notation
\be
\omega_{p}(1)=\xi (1)\psi_{0}\phi_{p}(1) ~~~.
\ee
For future reference it is not difficult to work out the
${\cal O}(2)$ contributions to
$Q_{4}$  and $\hat{Q}_{4}$ given by
\be
Q_{4}^{(2)}(1234)=
-\frac{\omega_{1}(1)}{2} \Bigg[G_{4}(1134)G_{2}(12)
+G_{4}(1124)G_{2}(13)
\nonumber
\ee
\be
+G_{4}(1123)G_{2}(14)\Bigg]
\ee
and
\be
\hat{Q}_{4}^{(2)}(1234)=
-\frac{\omega_{1}(1)}{2} \Bigg[G_{4}(1134)G_{Mm}(12)
+G_{4}(1124)G_{Mm}(13)
\nonumber
\ee
\be
+G_{4}(1123)G_{Mm}(14)\Bigg]
\nonumber
\ee
\be
-\omega_{1}(1)\Bigg[G_{Mmmmm}(1134)G_{2}(12)
+G_{Mmmmm}(1124)G_{2}(13)
\nonumber
\ee
\be
+G_{Mmmmm}(1123)G_{2}(14)\Bigg] ~~~.
\ee

Using
Eqs.(\ref{eq:204}) and (\ref{eq:82}), we easily find the determining
equation for $G_{Mmmm}$ at lowest nontrivial order:
\be
\int ~d\bar{1}G_{Mm}^{-1,0}(1\bar{1})G_{Mmmm}(\bar{1}234)
\nonumber
\ee
\be
+i\omega_{1}(1)\Bigg[G_{Mm}(12)G_{2}(13)G_{2}(14)
\ee
\be
+G_{Mm}(13)G_{2}(12)G_{2}(14)+G_{Mm}(14)G_{2}(12)G_{2}(13)
\Bigg] =0 ~~~.
\nonumber
\ee
This equation is easily integrated
to give
\be
G_{Mmmm}(1234)= \int ~d\bar{1} G_{Mm}^{(0)}(1\bar{1})[-i\omega_{1}(\bar{1})]
\label{eq:209}
\ee
\be
\times\Bigg[G_{Mm}(\bar{1}2)G_{2}(\bar{1}3)G_{2}(\bar{1}4)
+G_{Mm}(\bar{1}3)G_{2}(\bar{1}2)G_{2}(\bar{1}4)
\nonumber
\ee
\be
+G_{Mm}(\bar{1}4)G_{2}(\bar{1}2)G_{2}(\bar{1}3)\Bigg] ~~~.
\nonumber
\ee
Inserting Eq.(\ref{eq:203}) into Eq.(\ref{eq:83}), we have
\be
G_{mM}^{-1,0}(1\bar{1})G_{4}(\bar{1}234)
+i\omega_{1}(1)G_{2}(12)G_{2}(13)G_{2}(14)
\nonumber
\ee
\be
=-\Pi_{0}(1\bar{1})G_{Mmmm}(\bar{1}234)
\ee
where an integration over repeated barred indices here and below is assumed.
This can be integrated up to give
\be
G_{4}(1234)=
-G_{mM}^{(0)}(1\bar{1})\Pi_{0}(\bar{1}\bar{2})G_{Mmmm}(\bar{2}234)
\nonumber
\ee
\be
+G_{mM}^{(0)}(1\bar{1})
(-i\omega_{1}(\bar{1}))G_{2}(\bar{1}2)G_{2}(\bar{1}3)G_{2}(\bar{1}4) ~~~.
\ee
Inserting
the result for $G_{Mmmm}$ given by Eq.(\ref{eq:209}) and using
Eq.(\ref{eq:164}) gives the result
\be
G_{4}(1234)=G_{2}^{(0)}(1\bar{1})[-i\omega_{1}(\bar{1})]\Bigg[
G_{Mm}(\bar{1}2)G_{2}(\bar{1}3)G_{2}(\bar{1}4)
\nonumber
\ee
\be
+G_{Mm}(\bar{1}3)G_{2}(\bar{1}2)G_{2}(\bar{1}4)
+G_{Mm}(\bar{1}4)G_{2}(\bar{1}2)G_{2}(\bar{1}3)\Bigg]
\nonumber
\ee
\be
+G_{Mm}^{(0)}(\bar{1}1)
(-i\omega_{1}(\bar{1}))G_{2}(\bar{1}2)G_{2}(\bar{1}3)G_{2}(\bar{1}4) ~~~.
\label{eq:194}
\ee
If we evaluate all of the two-point correlations in $G_{4}$ and
$G_{Mmmm}$ at lowest order we obtain our lowest-order
approximation for
the $4$-point quantities.  Notice that $G_{4}$ is properly symmetric
under interchange of any two of its labels.

\section{Structure of Perturbation Theory at Higher Order}

In order to understand the structure of the perturbation theory one can
rather easily see that we should consider two classes of contributions to
\be
Q_{n}(12...n)=
\frac{\delta^{n-1}}{\delta h(n)\delta h(n-1)\cdots\delta h(2)}
\nonumber
\ee
\be
\times\left[
\int ~dx ~ \xi (1)\sigma_{1} (x)
\int\frac{dk}{2\pi}e^{-ikx}\Phi (k,h,1)
\right].
\ee
The first class corresponds to all $2n-1$
derivatives with respect to $h$ acting on
$\Phi (k,h,1)$ to give, in the zero external field limit,
a product of two-point
correlation functions of the form
\be
Q_{2n}(12\cdots ,2n)
\nonumber
\ee
\be
=\xi (1)\psi_{0}{\cal V}_{n-1}(1)
G_{2}(12)G_{2}(13)\cdots G_{2}(1,2n) ~~~.
\nonumber
\ee
There is another set of contributions where all of the derivatives
except the first acts on the factor multiplying $\Phi (k,h,1)$ and
gives a contribution
\be
Q_{2n}(12\cdots ,2n)=\xi (1)\psi_{0}{\cal V}_{0}(1)
G_{2n}(123\cdots ,2n) .
\nonumber
\ee
Since each  of these terms is of the same order in $\phi_{p}$, we
easily find the proposed result
$ G_{2p}\approx \phi_{p-1}$ ~.
A very similar analysis follows for the case of $\hat{Q}_{2n}$.

\section{Two-Point Correlation Function at Higher Order}

Following the same procedures it is easy to find that the next order
contribution to $Q_{2}$ and $\hat{Q}_{2}$ is of ${\cal O}(2)$ and given by
\be
Q_{2}^{(2)}(12)=
-\frac{\omega_{1}(1)}{3!}G_{4}(1112)
\ee
\be
\hat{Q}_{2}^{(2)}(12)=-
\frac{\omega_{1}(1)}{2}G_{Mmmm}(1112) ~~~.
\label{eq:218}
\ee
Inserting Eqs.(\ref{eq:153}) and (\ref{eq:218}) into Eq.(\ref{eq:90})
gives the equation for
the response function to second order:
\be
G_{Mm}^{-1,0}(1\bar{1})G_{Mm}(\bar{1}2)
+i\frac{\omega_{1}(1)}{2}G_{Mmmm}(1112)=\delta (12) .
\nonumber
\ee
Similarly, the equation determining the correlation function
to second order is given by
\be
G_{mM}^{-1,0}(1\bar{1})G_{2}(\bar{1}2)
+i\frac{\omega_{1}(1)}{3!}G_{4}(1112)
=-\Pi_{0}(1\bar{1})G_{Mm}(\bar{1}2) .
\nonumber
\ee
These equations can be integrated to give
\be
G_{Mm}(12)=G_{Mm}^{(0)}(12)
\nonumber
\ee
\be
+G_{Mm}^{(0)}(1\bar{1})
\frac{-i\omega_{1}(\bar{1})}{2}G_{Mmmm}(\bar{1}\bar{1}\bar{1}2)
\label{eq:221}
\ee
and
\be
G_{2}(12)=-G_{mM}^{(0)}(1\bar{1})
\Pi_{0}(\bar{1}\bar{2})G_{Mm}(\bar{2}2)
\nonumber
\ee
\be
+G_{mM}^{(0)}(1\bar{1})\frac{-i\omega_{\bar{1}}(1)}{3!}
G_{4}(\bar{1}\bar{1}\bar{1}2) ~~~.
\label{eq:222}
\ee
Inserting Eq.(\ref{eq:221}) for $G_{Mm}$ into  Eq.(\ref{eq:222})
gives
\be
G_{2}(12)=G_{2}^{(0)}(12)
+G_{2}^{(0)}(1\bar{1})\frac{-i\omega_{1}(\bar{1})}{2}
G_{Mmmm}(\bar{1}\bar{1}\bar{1}2)
\nonumber
\ee
\be
+G_{mM}^{(0)}(1\bar{1})\frac{-i\omega_{1}(\bar{1})}{3!}
G_{4}(\bar{1}\bar{1}\bar{1}2)  ~~~.
\ee
Using our ${\cal O}(1)$ result for
$G_{Mmmm}$ gives the ${\cal O}(2)$ result
for the response
function
\be
G_{Mm}(12)=G_{Mm}^{(0)}(12)
+G_{Mm}^{(0)}(1\bar{1})\Sigma_{Mm}^{(2)}(\bar{1}\bar{2})
G_{Mm}^{(0)}(\bar{2}2) ~~~,
\ee
where the lowest-order self-energy contribution is given by
\be
\Sigma_{Mm}^{(2)}(12)=\frac{1}{2}[-i\omega_{1}(1)]
G_{Mm}^{(0)}(12)G_{2}^{(0)}(12)^{2}[-i\omega_{1}(2)] ~~~.
\ee
Using the results for $G_{Mmmm}$ and $G_{4}$ at ${\cal O}(1)$
leads to the ${\cal O}(2)$ result for the
correlation function
\be
G_{2}(12)=G_{2}^{(0)}(12)+G_{2}^{(2,1)}(12)
\nonumber
\ee
\be
+G_{2}^{(2,1)}(21)
+G_{2}^{(2,2)}(12)
\nonumber
\ee
where
\be
G_{2}^{(2,1)}(12)=G_{2}^{(0)}(1\bar{1})
\Sigma_{Mm}^{(2)}(\bar{1}\bar{2})G_{Mm}^{(0)}(\bar{2}2)
\ee
and
\be
G_{2}^{(2,2)}(12)=
-G_{mM}^{(0)}(1\bar{1})\Pi^{(2)}(\bar{1}\bar{2})
G_{Mm}^{(0)}(\bar{2}2) ~~~.
\ee
The self-energy is the same as for the response function and
\be
\Pi^{(2)}(12)=-\frac{1}{3!}[-i\omega_{1}(1)]G_{2}^{(0)}(12)^{3}
[-i\omega_{1}(2)] ~~~.
\ee
We need to evaluate $G_{2}^{(2,1)}$ and $G_{2}^{(2,2)}$.
The
integrations over space, in $d$-dimensions,
are straightforward since they involve
products of displaced gaussians.  After rescaling the internal
time integrations $\bar{t}_{1}=Ty_{1}$, $\bar{t}_{2}=Ty_{2}$,
$T=\frac{1}{2}(t_{1}+t_{2})$, we
obtain
\be
G_{2}^{(2,1)}(12)=\sqrt{S_{0}(1)S_{0}(2)}2^{d-1}
\omega^{2}\Phi_{0}(t_{1}t_{2})
J_{1}(x,t_{1}/T,T)
\nonumber
\ee
\be
J_{1}(x,t_{1}/T,T)=
\int_{t_{0}/T}^{t_{1}/T}dy_{1}\int_{t_{0}/T}^{y_{1}}dy_{2}
R_{1}(y_{1},y_{2})
e^{-\frac{1}{2}g_{1}(y_{1},y_{2})x^{2}}
\nonumber
\ee
\be
R_{1}(y_{1},y_{2})=
\frac{y_{1}^{d/2-1}y_{2}^{d/2-1}}
{[(y_{1}+y_{2})(3y_{1}-y_{2}-(y_{1}-y_{2})^{2})]^{d/2}}
\nonumber
\ee
\be
G_{2}^{(2,2)}(12)=\sqrt{S_{0}(1)S_{0}(2)}\frac{2^{d-1}}{3}
\omega^{2}\Phi_{0}(t_{1}t_{2})
J_{2}(x,t_{1},t_{2})
\nonumber
\ee
\be
J_{2}(x,t_{1},t_{2})=
\int_{t_{0}/T}^{t_{1}/T}dy_{1}\int_{t_{0}/T}^{t_{2}/T}dy_{2}
R_{2}(y_{1},y_{2})
e^{-\frac{1}{2}g_{2}(y_{1},y_{2})x^{2}}
\nonumber
\ee
\be
R_{2}(y_{1},y_{2})=
\frac{y_{1}^{d/2-1}y_{2}^{d/2-1}}
{[(y_{1}+y_{2})^{2}(3-y_{1}-y_{2})]^{d/2}}
\nonumber
\ee
where we have chosen $x^{2}=r^{2}/4T$ and
\be
g_{1}(y_{1},y_{2})=\frac{3y_{1}-y_{2}}{3y_{1}-y_{2}-(y_{1}-y_{2})^{2}}
\nonumber
\ee
and
\be
g_{2}(y_{1},y_{2})=\frac{3}{3-y_{1}-y_{2}} ~~~.
\nonumber
\ee

The first thing we should do with this result is look at the contribution
at this order to the onsite equal-time $t_{1}=t_{2}=t$
correlation function given by
\be
S_{2}(t) =S_{2}^{(0)}(t)+2G_{2}^{(2,1)}(11)+G_{2}^{(2,2)}(11)
\nonumber
\ee
\be
=S_{2}^{(0)}(t)\left[1 +2^{d}\omega^{2}J_{1}(0,1,t)+
\frac{2^{d}}{6}\omega^{2}J_{2}(0,t,t)\right]
\nonumber
\ee
where we have the integrals
\be
J_{1}(0,1,t)=
\int_{t_{0}/t}^{1}dy_{1}\int_{t_{0}/t}^{y_{1}}dy_{2}R_{1}(y_{1},y_{2})
\nonumber
\ee
\be
J_{2}(0,t,t)=
\int_{t_{0}/t}^{1}dy_{1}\int_{t_{0}/t}^{1}dy_{2}R_{2}(y_{1},y_{2}) ~~~.
\nonumber
\ee
The key point here is that $J_{1}(0,1,t)$ and $J_{2}(0,t,t)$ are
logarithmically divergent at $t\rightarrow\infty$.
One can show
to logarithmic order that
\be
J_{1}(0,1,t)=K_{d}~ln(t/t_{0})+\cdots
\nonumber
\ee
where
\be
K_{d}=\int_{0}^{1}dz \frac{z^{d/2-1}}{[(1+z)(3-z)]^{d/2}}
\nonumber
\ee
and
\be
J_{2}(0,t,t)=\frac{2}{3^{d/2}}M_{d}~ln(t/t_{0})+\cdots
\nonumber
\ee
where
\be
M_{d}=\int_{0}^{1}dz \frac{z^{d/2-1}}{[1+z]^{d}}=
\frac{1}{2}\frac{\Gamma^{2}(d/2)}{\Gamma (d)} ~~~.
\nonumber
\ee
We have then that
\be
S_{2}(1)=S_{2}^{(0)}(t)\left[1+\omega^{2}2^{d}\left(K_{d}+
\frac{M_{d}}{3^{d/2+1}}\right) ln(t/t_{0})+\cdots \right]
\nonumber
\ee
and a simple exponentiation of this result gives
\be
S_{2}(1)=S_{2}^{(0)}(t)
\left(\frac{t}{t_{0}}\right)^{\omega^{2}2^{d}\left(K_{d}+
\frac{M_{d}}{3^{d/2+1}}\right)}\left( 1 + \cdots\right) ~~~.
\nonumber
\ee
For self-consistency we must determine $\omega$ at this order.
Remembering that at lowest order
\be
S_{2}^{(0)}(t)=A_{0}t^{2\omega -d/2},
\nonumber
\ee
we require that
\be
S_{2}(1)=
A_{0}\left(\frac{t}{t_{0}}\right)^{2\omega -d/2+\omega^{2}2^{d}\left(K_{d}+
\frac{M_{d}}{3^{d/2+1}}\right)}\left( 1 + \cdots\right)=At
\nonumber
\ee
which determines  $\omega$ at this order
\be
2\omega+\omega^{2}2^{d}\left(K_{d}+
\frac{M_{d}}{3^{d/2+1}}\right)=1+\frac{d}{2} ~~~.
\ee

Then, for example, for $d=2$
\be
K_{2}=\frac{1}{4}ln~3  ~~~,~~~M_{2}=\frac{1}{2}
\nonumber
\ee
and
\be
\omega =\frac{\sqrt{1+2~ln~ 3+4/9}-1}{ln~3 +2/9}
=0.687687370 \ldots
\nonumber
\ee
For large d analytical progress leads to the results
\be
K_{d}\approx \frac{2}{d}\frac{1}{2^{d}}+\cdots
\nonumber
\ee
\be
M_{d}\approx \sqrt{\frac{2\pi}{d}}\frac{1}{2^{d}}+\cdots
\nonumber
\ee
and
\be
\omega =\frac{d}{2}\left(\sqrt{2}-1\right) ~~~.
\nonumber
\ee
A numerical determination of
$\omega$ shows that it is approximately linear with
d over the  whole range of d.

The $t/t_{0}\rightarrow \infty $ singularity in $G_{2}(12)$ at ${\cal O}(2)$
can be regulated by turning our
attention from $G_{2}(12)$ to the quantities
$\Phi (t_{1}t_{2})$ and $f(x,t_{1}/t_{2})$ defined by
\be
G_{2}(12)=\sqrt{S(1)S(2)}\Phi (t_{1}t_{2})f(x,t_{1}/t_{2})
\ee
and the constraint
$f(x=0,t_{1}/t_{2})=1$~.
If we write
\be
G_{2}(12)=G_{2}^{(0)}(12)+\Delta G_{2}(12)
\nonumber
\ee
and
\be
S_{2}(1)=S_{2}^{(0)}(1)+\Delta S_{2}(1) ~~~,
\nonumber
\ee
then
\be
\Phi (t_{1}t_{2})f(x,t_{1}/t_{2})=
\frac{G_{2}(12)}{\sqrt{S_{2}(1)S_{2}(2)}}
\ee
\be
=\Phi ^{(0)}(t_{1},t_{2})f_{0}(x)
\left(1-\frac{1}{2}\left[
\frac{\Delta S_{2}(1)}{S_{2}^{(0)}(1)}
+\frac{\Delta S_{2}(2)}{S_{2}^{(0)}(2)}\right]\right)
\nonumber
\ee
\be
+\frac{\Delta G_{2}(12)}{\sqrt{S_{2}^{(0)}(1)S_{2}^{(0)}(2)}}
{}~~~ .
\nonumber
\ee
We can then separate the contribution to the on-site correlation
function $\Phi (t_{1},t_{2})$ from the general $x$-dependence and write
\be
\Phi (t_{1},t_{2})=\Phi_{0}(t_{1},t_{2})\left( 1 +\omega^{2}
2^{d}\Delta\Phi (t_{1},t_{2})\right)
\ee
\be
f(x,t_{1}/t_{2})=f_{0}(x)\left( 1 +\omega^{2}
2^{d}W(x, t_{1}/t_{2})\right)
\ee
where
\be
\Delta \Phi=\frac{1}{2}\Phi_{1}+\frac{1}{6}\Phi_{2}
\ee
\be
\Phi_{1}=
\frac{1}{2}\int_{t_{0}/T}^{t_{1}/T}~dy_{1}
\int_{t_{0}/T}^{y_{1}}~dy_{2}R_{1}(y_{1},y_{2})
\nonumber
\ee
\be
+\frac{1}{2}\int_{t_{0}/T}^{t_{2}/T}~dy_{1}
\int_{t_{0}/T}^{y_{1}}~dy_{2}R_{1}(y_{1},y_{2})
\nonumber
\ee
\be
-\frac{1}{2}\int_{t_{0}/t_{1}}^{1}~dy_{1}
\int_{t_{0}/t_{1}}^{y_{1}}~dy_{2}R_{1}(y_{1},y_{2})
\nonumber
\ee
\be
-\frac{1}{2}\int_{t_{0}/t_{2}}^{1}~dy_{1}
\int_{t_{0}/t_{2}}^{y_{1}}~dy_{2}R_{1}(y_{1},y_{2})
\nonumber
\ee
\be
\Phi_{2}=
\frac{1}{6}\int_{t_{0}/T}^{t_{1}/T}~dy_{1}
\int_{t_{0}/T}^{t_{2}/T}~dy_{2}R_{2}(y_{1},y_{2})
\nonumber
\ee
\be
-\frac{1}{12}\int_{t_{0}/t_{1}}^{1}~dy_{1}
\int_{t_{0}/t_{1}}^{1}~dy_{2}R_{2}(y_{1},y_{2})
\nonumber
\ee
\be
-\frac{1}{12}\int_{t_{0}/t_{2}}^{1}~dy_{1}
\int_{t_{0}/t_{2}}^{1}~dy_{2}R_{2}(y_{1},y_{2})
\nonumber
\ee
and
\be
W(x,t_{1}/t_{2})=
\frac{1}{2}\int_{0}^{t_{1}/T}dy_{1}
\int_{0}^{y_{1}}~dy_{2}R_{1}(y_{1},y_{2})
\left[e^{-\frac{1}{2}\Delta g_{1}x^{2}}-1\right]
\nonumber
\ee
\be
+\frac{1}{2}\int_{0}^{t_{2}/T}~dy_{1}
\int_{0}^{y_{1}}~dy_{2}R_{1}(y_{1},y_{2})
\left[e^{-\frac{1}{2}\Delta g_{1}x^{2}}-1\right]
\nonumber
\ee
\be
+\frac{1}{6}\int_{0}^{t_{1}/T}~dy_{1}
\int_{0}^{t_{2}/T}~dy_{2}R_{2}(y_{1},y_{2})
\left[e^{-\frac{1}{2}\Delta g_{2}x^{2}}-1\right]
\nonumber
\ee
where
\be
\Delta g_{1}=g_{1}(y_{1},y_{2})-1=\frac{(y_{1}-y_{2})^{2}}
{3y_{1}-y_{2}-(y_{1}-y_{2})^{2}}
\nonumber
\ee
and
\be
\Delta g_{2}=g_{2}(y_{1},y_{2})-1=\frac{y_{1}+y_{2}}{3-y_{1}-y_{2}} ~~~.
\nonumber
\ee

Let us first look first at $\Delta\Phi (t_{1},t_{2})$ and the integrals
$\Phi_{1}$ and $\Phi_{2}$.  An investigation of $\Phi_{1}$ shows that it is a
regular quantity for $t_{1}\gg t_{2}$ or $t_{2}\gg t_{1}$.  Turning to
$\Phi_{2}$, however, one finds that it is logarithmically divergent in these
limits:
\be
\Phi_{2}=\frac{2M_{d}}{3^{d/2}}~ln~\left(\frac{\sqrt{t_{1}t_{2}}}{T}\right)
+\cdots
\nonumber
\ee
This means to second order we have
\be
\Phi (t_{1},t_{2})=\left(\frac{\sqrt{t_{1}t_{2}}}{T}\right)^{d/2}
\left( 1 +\omega^{2}
\frac{2^{d}M_{d}}{3^{d/2}+1}~ln~\left(\frac{\sqrt{t_{1}t_{2}}}{T}\right)
\right)
\nonumber
\ee
which can be exponentiated to give
\be
\Phi (t_{1},t_{2})=\left(\frac{\sqrt{t_{1}t_{2}}}{T}\right)^{\lambda_{m}}
\left( 1 +\cdots\right)
\ee
where
\be
\lambda_{m} =\frac{d}{2}+\omega^{2}\frac{2^{d}M_{d}}{3^{d/2+1}}
{}~~~.
\ee
It is just this expression for $\lambda$ discussed in Section II.

Turning next to $f(x,t_{1}/t_{2})$, one of the first things to note is that
it is for small $x$.  Looking at the first term in the power
series expansion in $x^{2}$ we obtain
\be
W(x,t_{1}/t_{2})=-\frac{1}{2}x^{2}\Bigg[\frac{1}{2}I_{1}^{(2)}(t_{1}/T)
\nonumber
\ee
\be
+\frac{1}{2}I_{1}^{(2)}(t_{2}/T)+\frac{1}{6}I_{2}^{(2)}(t_{1}/t_{2})\Bigg]
\nonumber
\ee
where
\be
I_{1}^{(2)}(t_{1}/T)=
\int_{0}^{t_{1}/T}~dy_{1}
\int_{0}^{y_{1}}~dy_{2}R_{1}(y_{1},y_{2})
\Delta g_{1}(y_{1},y_{2})
\nonumber
\ee
\be
I_{2}^{(2)}(t_{1}/t_{2})=
\int_{0}^{t_{1}/T}~dy_{1}
\int_{0}^{t_{2}/T}~dy_{2}R_{2}(y_{1},y_{2})
\Delta g_{2}(y_{1},y_{2})
\nonumber
\ee
where since $T\gg t_{0}$, the lower limits can be set to zero.  After
making the change of variables $y_{2}=y_{1}z$ one finds that one can
perform the integral over $y_{1}$ to obtain
\be
I_{1}^{(2)}(t_{1}/T)=\frac{2}{d}\left[J_{d}(t_{1}/T)-K_{d}(t_{1}/T)\right]
\ee
where
\be
J_{d}(t_{1}/T)=\int_{0}^{t_{1}/T}dz \frac{z^{d/2-1}}{[(1+z)^{2}(2-z)]^{d/2}}
\ee
\be
K_{d}(t_{1}/T)=\int_{0}^{t_{1}/T}dz \frac{z^{d/2-1}}{[(1+z)(3-z)]^{d/2}}
\ee
\be
I_{2}^{(2)}(t_{1}/T)=\frac{2}{d}\left[L_{d}(t_{1}/t_{2})
+L_{d}(t_{1}/t_{2})-\frac{2}{3^{d/2}}M_{d}\right]
\ee
and
\be
L_{d}(t_{1}/t_{2})=\int_{0}^{t_{1}/t_{2}}dz \frac{z^{d/2-1}}
{[(1+z)^{2}(3-\frac{t_{2}}{T}(1+z))]^{d/2}} ~~~.
\label{eq:281}
\ee
Notice that at equal times, $K_{d}(1)=K_{d}$, $L_{d}(1)=J_{d}(1)$,
and all of these integrals are well behaved.

There is one last regularization that must be carried out before the
perturbation theory expression for the correlation function can be
used for all values of the parameters.  Consider that $W$
can be written as the sum of three terms:
\be
W^{(1)}(x,t_{1}/t_{2})=
\frac{1}{2}\int_{0}^{t_{1}/T}~dy_{1}
\int_{0}^{y_{1}}~dy_{2}R_{1}(y_{1},y_{2})
\nonumber
\ee
\be
\times\left[e^{-\frac{1}{2}\Delta g_{1}x^{2}}-1\right]
\nonumber
\ee
\be
W^{(1)}(x,t_{2}/t_{1})=
\frac{1}{2}\int_{0}^{t_{2}/T}~dy_{1}
\int_{0}^{y_{1}}~dy_{2}R_{1}(y_{1},y_{2})
\nonumber
\ee
\be
\times\left[e^{-\frac{1}{2}\Delta g_{1}x^{2}}-1\right]
\nonumber
\ee
\be
W^{(2)}(x,t_{1}/t_{2})=
\frac{1}{6}\int_{0}^{t_{1}/T}~dy_{1}
\int_{0}^{t_{2}/T}~dy_{2}R_{2}(y_{1},y_{2})
\nonumber
\ee
\be
\times\left[e^{-\frac{1}{2}\Delta g_{2}x^{2}}-1\right] ~~~.
\nonumber
\ee
In $W^{(1)}(x,t_{1}/t_{2})$ let $y_{2}=y_{1}z$ which leads to the
result
\be
W^{(1)}(x,t_{1}/t_{2})=
\frac{1}{2}\int_{0}^{t_{1}/T}\frac{dy_{1}}{y_{1}}
\nonumber
\ee
\be
\times\int_{0}^{1}~dz \frac{z^{d/2-1}}{[(1+z)(3-z-y_{1}(1-z)^{2})]^{d/2}}
\nonumber
\ee
\be
\times\left[e^{-\frac{1}{2}x^{2}\frac{y_{1}(1-z)^{2}}{(3-z-y_{1}(1-z)^{2})}}-1\right]
{}~~~.
\ee
Notice in this integral that there is an apparent log divergence for small
$y_{1}$ which is cancelled between the exponential term and the subtraction
term.  Note that the log terms survives if $x$ is large enough.  To pick
out the log term we can expand about small $y_{1}$ except for
the contribution $x^{2}y_{1}$
and write
\be
W^{(1)}(x,t_{1}/t_{2})=
\frac{1}{2}\int_{0}^{t_{1}/T}\frac{dy_{1}}{y_{1}}
\int_{0}^{1}~dz \frac{z^{d/2-1}}{[(1+z)(3-z)]^{d/2}}
\nonumber
\ee
\be
\times\left[e^{-\frac{1}{2}x^{2}\frac{y_{1}(1-z)^{2}}{(3-z)}}-1\right]
+\Delta W^{(1)}(x,t_{1}/t_{2})
\nonumber
\ee
where
$\Delta W^{(1)}(x,t_{1}/t_{2})$ has no singularity for large
$x$.  If we now make the change of variables
\be
y_{1}=\frac{t_{1}s}{T(1+x^{2})}
\ee
in the leading integral we obtain
\be
W^{(1)}(x,t_{1}/t_{2})=
\frac{1}{2}\int_{0}^{1+x^{2}}\frac{ds}{s}
\nonumber
\ee
\be
\times\int_{0}^{1}~dz \frac{z^{d/2-1}}{[(1+z)(3-z)]^{d/2}}
\left[e^{-\frac{1}{2}\frac{x^{2}}{1+x^{2}}\frac{t_{1}}{T}
\frac{s(1-z)^{2}}{(3-z)}}-1\right] +\cdots
\nonumber
\ee
where the $\cdots$ refer to contributions which are regular.  The integral
over $s$ can be divided into a regular part from $0$ to $1$ and the
singular part for large $x$ given by
\be
W^{(1)}(x,t_{1}/t_{2})=
\frac{1}{2}\int_{1}^{1+x^{2}}\frac{ds}{s}
\int_{0}^{1}~dz \frac{z^{d/2-1}}{[(1+z)(3-z)]^{d/2}}
\nonumber
\ee
\be
\times\left[e^{-\frac{1}{2}\frac{x^{2}}{1+x^{2}}\frac{t_{1}}{T}
\frac{s(1-z)^{2}}{(3-z)}}-1\right] +\cdots
\nonumber
\ee
The singular part is now isolated  in the second piece that does not
have  exponential convergence for large $s$ .  We have then
\be
W^{(1)}(x,t_{1}/t_{2})=-\frac{1}{2}~K_{d}~ln(1+x^{2})+\cdots
\ee
Clearly
\be
W^{(1)}(x,t_{2}/t_{1})=-\frac{1}{2}~K_{d}~ln(1+x^{2})+\cdots
\ee
Turning to the third contribution we again isolate the small $y_{1}$
and $y_{2}$ behaviors and expanding in $x^{2}$ in places which do not
contribute to the singularity gives
\be
W^{(2)}(x,t_{1}/t_{2})=
\frac{1}{6}\int_{0}^{t_{1}/T}~dy_{1}
\int_{0}^{t_{2}/T}~dy_{2}\frac{y_{1}^{d/2-1}y_{2}^{d/2-1}}
{(y_{1}+y_{2})^{d}}\frac{1}{3^{d/2}}
\nonumber
\ee
\be
\times\left[e^{-\frac{1}{6}x^{2}(y_{1}+y_{2})}-1\right]+\cdots
\nonumber
\ee
Next make the coordinate transformations
\be
y_{1}=\frac{s_{1}t_{1}}{T(1+x^{2})}
\ee
\be
y_{2}=\frac{s_{2}t_{2}}{T(1+x^{2})}
\ee
which results in the leading contribution to the integral
\be
W^{(2)}(x,t_{1}/t_{2})=
\frac{1}{2~3^{d/2+1}}\int_{0}^{1+x^{2}}ds_{1}
\int_{0}^{1+x^{2}}ds_{2}
\nonumber
\ee
\be
\times\frac{s_{1}^{d/2-1}s_{2}^{d/2-1}}
{(s_{1}+s_{2})^{d}}
\Bigg[e^{-\frac{1}{6}\frac{x^{2}}{1+x^{2}}
(\frac{s_{1}t_{1}}{T}+\frac{s_{2}t_{2}}{T}))}-1\Bigg]+\cdots
\nonumber
\ee
Again the leading behavior for large $x$ comes from the terms which do
not have exponential convergence for large $s_{1}$ and $s_{2}$:
\be
W^{(2)}(x,t_{1}/t_{2})=
-\frac{1}{2~3^{d/2+1}}\int_{0}^{1+x^{2}}ds_{1}
\int_{0}^{1+x^{2}}ds_{2}
\nonumber
\ee
\be
\times\frac{s_{1}^{d/2-1}s_{2}^{d/2-1}}
{(s_{1}+s_{2})^{d}}
+\cdots
\nonumber
\ee
If we let $s_{1}\rightarrow 1/s_{1}$ and $s_{2}\rightarrow 1/s_{2}$
then this integral takes the form
\be
W^{(2)}(x,t_{1}/t_{2})=
-\frac{1}{2~3^{d/2+1}}\int_{1/(1+x^{2})}^{1}ds_{1}
\int_{1/(1+x^{2})}^{1}ds_{2}\nonumber
\ee
\be
\times\frac{s_{1}^{d/2-1}s_{2}^{d/2-1}}
{(s_{1}+s_{2})^{d}}
+\cdots
\nonumber
\ee
This integral was evaluated in our treatment of $\Phi$ with the result
\be
W^{(2)}(x,t_{1}/t_{2})=
-\frac{1}{2~3^{d/2+1}} 2M_{d}~ln~\left[\frac{1+x^{2}}{2}\right]
+\cdots
\nonumber
\ee
Then to leading order for large $x$ we have
\be
W(x,t_{1}/t_{2})= -K_{d}~ln(1+x^{2})
-\frac{1}{3^{d/2+1}} M_{d}~ln\left[\frac{1+x^{2}}{2}\right]
+\cdots
\nonumber
\ee
The scaled correlation function then has the form
\be
f(x,t_{1}/t_{2})=f_{0}(x)\left[1-\omega^{2}2^{d+1}\left(
K_{d}+\frac{M_{d}}{3^{d/2+1}}\right)~ln~(1+x^{2})\right]
\nonumber
\ee
\be
\times\left[1+\cdots\right] ~~~.
\ee
This can be exponentiated to obtain
\be
f(x,t_{1}/t_{2})=\frac{f_{0}(x)}{(1+x^{2})^{\nu_{m} /2}}\left[1+\cdots\right]
\ee
where the exponent governing the large $x$ behavior is given by
\be
\nu_{m} =\omega^{2}2^{d+1}\left(
K_{d}+\frac{M_{d}}{3^{d/2+1}}\right) ~~~.
\ee

\section{The Order Parameter Correlation Function}

\subsection{Perturbation Expansion}

We turn next to the connection between the correlation function for the
auxiliary field $m$ and the order parameter correlation function.  If we
look at the problem using the transformation given by Eq.(\ref{eq:30b}),
we have
\be
C(12)=\left<\psi (1)\psi (2)\right>
\nonumber
\ee
\be
=\left<\sigma (1)\sigma (2)\right>+\left<\sigma (1)u(2)\right>+
\left<u(1)\sigma (2)\right>+\left<u(1)u(2)\right> ~~~.
\nonumber
\ee
The key point is that since $u(1)$ vanishes exponentially for large
$|m(1)|$, the averages over these fields are down by a factor of
$L^{-2}$ relative to the averages over the field $\sigma (1)$.  Thus
$\left<\sigma (1)\sigma (2)\right>\approx{\cal O}(1)$ as
$L(t)\rightarrow\infty$, while $\left<\sigma (1)u(2)\right>$
and $\left<u(1)\sigma (2)\right>$ are of ${\cal O}(L^{-2})$
and $\left<u(1)u(2)\right>$ of ${\cal O}(L^{-4})$.
In the scaling regime we have
\be
C(12)=\left<\sigma (1)\sigma (2)\right> ~~~.
\ee
This quantity can evaluated using the two-point probability distribution
\be
P_{h}(x_{1}x_{2},1,2)=<\delta (x_{1}-m(1))\delta (x_{2}-m(2))>_{h}
\ee
and the correlation functions are obtained at zero external field via:
\be
C(12)=\int~dx_{1}\int~dx_{2}~\sigma (x_{1})\sigma (x_{2})
P_{0}(x_{1}x_{2},12) ~~~.
\ee
More generally we can treat the set of correlation functions
\be
C_{n\ell}(12)=\left<\sigma_{n} (1)\sigma_{\ell} (2)\right>
\nonumber
\ee
\be
=\int~dx_{1}\int~dx_{2}~\sigma_{n} (x_{1})\sigma_{\ell} (x_{2})
P_{0}(x_{1}x_{2},12) ~~~.
\nonumber
\ee
Since we have computed the auxiliary field correlation functions
to ${\cal O}(2)$ we also need to determine $C_{n\ell}(12)$
to second order.  This
expansion can be
developed as follows. As in the case of the one-point quantity
$P_{h}(x_{1},1)$,  we again
use the
integral representation for the $\delta$-function to obtain
\be
P_{h}(x_{1},x_{2},12)= \int\frac{dk_{1}}{2\pi}
\int\frac{dk_{2}}{2\pi}
e^{-ik_{1}x_{1}}e^{-ik_{2}x_{2}}
<e^{{\cal H}(12)}>_{h}
\nonumber
\ee
where
${\cal H}(12)\equiv ik_{1}m(1)+ik_{2}m(2)$ ~.
The average of the exponential is precisely of the form which can be
rewritten in terms of cumulants:
\be
<e^{{\cal H}(12)}>_{h}=exp\left[\sum_{n=1}^{\infty}\frac{1}{n!}G_{{\cal H}}^{(n)}(12)\right]
\ee
where
$G_{{\cal H}}^{(n)}(12)$ is the $n^{th}$ order cumulant
for the field ${\cal H}(12)$.  These cumulants can all be expressed
in terms of the $m$-field cumulants. In this section we can work
directly in terms of zero external fields where all odd $m$-field
cumulants vanish.  We will need here
\be
G_{{\cal H}}^{(2)}(12)=(ik_{1})^{2}G_{2}(11)+(ik_{2})^{2}G_{2}(22)
+2(ik_{1})(ik_{2})G_{2}(12)
\nonumber
\ee
\be
G_{{\cal H}}^{(4)}(12)=(ik_{1})^{4}G_{4}(1111)
+4(ik_{1})^{3}(ik_{1})G_{4}(1112)
\nonumber
\ee
\be
+6(ik_{1})^{2}(ik_{2})^{2}G_{4}(1122)+4(ik_{1})(ik_{2})^{3}G_{4}(1222)
\nonumber
\ee
\be
+(ik_{2})^{4}G_{4}(2222)
\nonumber
\ee
and
\be
G_{{\cal H}}^{(6)}(12)=(ik_{1})^{6}G_{6}(111111)
+6(ik_{1})^{5}(ik_{2})G_{6}(111112)
\nonumber
\ee
\be
+15(ik_{1})^{4}(ik_{2})^{2}G_{6}(111122)
\nonumber
\ee
\be
+20(ik_{1})^{3}(ik_{2})^{3}G_{6}(111222)
+15(ik_{1})^{2}(ik_{2})^{4}G_{6}(112222)
\nonumber
\ee
\be
+6(ik_{1})(ik_{2})^{5}G_{6}(122222)+(ik_{2})^{6}G_{6}(222222) ~~~.
\nonumber
\ee
Working to second order in the $\phi$-expansion requires
keeping terms
\be
<e^{{\cal H}(12)}>=\Phi_{0}(k_{1},k_{2},12)\Biggl(1
+\frac{1}{4!}G_{{\cal H}}^{(4)}
+\frac{1}{2}\left(\frac{1}{4!}G_{{\cal H}}^{(4)}\right)^{2}
\nonumber
\ee
\be
+\frac{1}{6!}G_{{\cal H}}^{(6)}\cdots\Biggr)
\label{eq:309}
\ee
where in zero external field $\Phi_{0}(k_{1},k_{2},12)$ is given
by
\be
\Phi_{0}(k_{1}k_{2},12)=
e^{-\frac{1}{2}\left[k_{1}^{2}G_{2}(11) +k_{2}^{2}G_{2}(22)
+2k_{1}k_{2}G_{2}(12)\right]} ~~~.
\ee

\subsection{Terms of ${\cal O}(1)$}

Let us look first at the theory keeping terms up to
${\cal O}(1)$.  It is then straighforward to show that the
two-point probability distribution is given by
\be
P_{0}(x_{1},x_{2};12)=\Biggl[1+\frac{1}{4!}\Biggl(G_{4}(1111)
\frac{d^{4}}{dx_{1}^{4}}
\nonumber
\ee
\be
+4G_{4}(1112)
\frac{d^{3}}{dx_{1}^{3}}\frac{d}{dx_{2}}
\nonumber
\ee
\be
+6G_{4}(1122)\frac{d^{2}}{dx_{1}^{2}}\frac{d^{2}}{dx_{2}^{2}}
+4G_{4}(1222)\frac{d}{dx_{1}}\frac{d^{3}}{dx_{2}^{3}}
\nonumber
\ee
\be
+G_{4}(2222)\frac{d^{4}}{dx_{2}^{4}}\Biggr)\Biggr]
P_{0}^{(0)}(x_{1},x_{2};12)
\label{eq:311}
\ee
where $P_{0}^{(0)}(x_{1},x_{2};12)$ is given by
\be
P_{0}^{(0)}(x_{1}x_{2},12)
=\frac{1}{2\pi}\frac{\gamma (12)}{\sqrt{S_{2}(1)S_{2}(2)}}
\nonumber
\ee
\be
\times e^{\left[-\frac{\gamma^{2} (12)}{2S_{2}(1)S_{2}(2)}
\left[x_{1}^{2}S_{2}(2)+x_{2}^{2}S_{2}(1)
-2G_{2}(12)x_{1}x_{2}\right]\right]}
\ee
\be
\gamma (12)=\frac{1}{\sqrt{1-f^{2}(12)}}
\ee
and
\be
f(12)=\frac{G_{2}(12)}{\sqrt{S_{2}(1)S_{2}(2)}} ~~~.
\ee
Notice that we already have one resummation here since it is the full
$G_{2}$ which appears in $P_{0}^{(0)}(x_{1}x_{2},12)$.  The general
set of two-point order-parameter correlation functions are given,
up to ${\cal O}(1)$ by
\be
C_{n\ell}(12)=C_{n\ell}^{(0)}(12)+C_{n\ell}^{(1)}(12)
\cdots
\ee
where the first-order correction to the leading gaussian behavior is
given explicitly by
\be
C_{n\ell}^{(1)}(12)=\frac{1}{4!}G_{4}^{(1)}(1111)C_{n+4,\ell}^{(0)}(12)
\nonumber
\ee
\be
+\frac{1}{3!}G_{4}^{(1)}(1112)C_{n+3,\ell +1}^{(0)}(12)
\nonumber
\ee
\be
+\frac{1}{4}G_{4}^{(1)}(1122)C_{n+2,\ell +2}^{(0)}(12)
+\frac{1}{3!}G_{4}^{(1)}(1222)C_{n+1,\ell +3}^{(0)}(12)
\nonumber
\ee
\be
+\frac{1}{4!}G_{4}^{(1)}(2222)C_{n,\ell +4}^{(0)}(12)
\nonumber
\ee
where we remember that $C_{n\ell}^{(0)}(12)$ is a known functional
of the exact $G_{2}(12)$ and the $G_{4}$'s are also the exact
quantities.  Thus $C_{n\ell}^{(0)}(12)$ and $C_{n\ell}^{(1)}(12)$
both contain contributions of ${\cal O}(2)$.

Thus determining the $C_{n\ell}$ to first order requires first
determining $G_{2}$ to first order, then evaluating the
$C_{n\ell}^{(0)}(12)$ as functions of $G_{2}$, and finally the
evaluation of the various two-point contractions of
$G_{4}(1234)$ evaluated at first order.
We have already evaluated $G_{2}^{(0)}$ and we have
the appropriate first-order expression for $G_{4}(1234)$.  Notice
that it is the same set of contracted four-point quantities which
enter into the determination of any set $n,\ell$.  Let us restrict our
subsequent analysis to the correlation functions
$C_{00}(12)$.  The first-order correction to
the leading gaussian result
\be
C_{00}^{(0)}(12) =\frac{2}{\pi}\psi_{0}^{2}sin^{-1}f(12)
\label{eq:D25}
\ee
is given by
\be
C_{00}^{(1)}(12)=\frac{1}{4!}G_{4}^{(1)}(1111)C_{4,0}^{(0)}(12)
+\frac{1}{3!}G_{4}^{(1)}(1112)C_{3,1}^{(0)}(12)
\nonumber
\ee
\be
+\frac{1}{4}G_{4}^{(1)}(1122)C_{2,2}^{(0)}(12)
+\frac{1}{3!}G_{4}^{(1)}(1222)C_{1,3}^{(0)}(12)
\nonumber
\ee
\be
+\frac{1}{4!}G_{4}^{(1)}(2222)C_{0,4}^{(0)}(12) ~~~.
\label{eq:297}
\ee

The zeroth-order correlation functions $C_{n,\ell}^{(0)}(12)$ can be
evaluated using the identities
\be
C_{n+1,\ell +1}^{(0)}(12)=\frac{\partial}{\partial G_{2}(12)}
C_{n,\ell}^{(0)}(12)
\ee
and
\be
C_{n+2,\ell }^{(0)}(12)=2\frac{\partial}{\partial S_{2}(1) }
C_{n,\ell}^{(0)}(12)
\label{eq:299}
\ee
derived in TUG.
Starting with the expression for $C_{0,0}^{(0)}(12)$
given by Eq.(\ref{eq:D25}), all of the other quantities can be calculated by
taking derivatives.  A summary of the results we need for
$C_{0,0}^{(1)}(12)$ is given by
\be
C_{40}^{(0)}(12) =\frac{2}{\pi}\frac{\psi_{0}^{2}}{S_{2}^{2}(1)}
\gamma^{3}f(3-2f^{2})
\ee
\be
C_{31}^{(0)}(12) =-\frac{2}{\pi}
\frac{\psi_{0}^{2}}{S_{2}^{3/2}(1)S_{2}^{1/2}(2)}
\gamma^{3}
\ee
\be
C_{22}^{(0)}(12) =\frac{2}{\pi}
\frac{\psi_{0}^{2}}{S_{2}(1)S_{2}(2)}
\gamma^{3} f ~~~.
\label{eq:302}
\ee

Next we need the various contracted four-point cumulants appearing
in Eq.(\ref{eq:297}).  These can all be evaluated using
Eq/(\ref{eq:194}).  First we have the fully contracted four-point
cumulant:
\be
G_{4}^{(0)}(1111)=-4\omega 2^{d}S_{2}^{2}(1)L_{d}
\label{eq:303}
\ee
where
\be
L_{d}=\int_{0}^{1}dy\frac{y^{d/2-1}}{\left[(2-y)(1+y)^{2}\right]^{d/2}}
\ee
is the equal-time limit of $L_{d}(t_{1}/t_{2})$ defined by
Eq.(\ref{eq:281}).
Next we have
\be
G_{4}^{(0)}(1122)=-2\omega 2^{d}S_{2}(1)S_{2}(2)
\nonumber
\ee
\be
\times\left[\xi^{d/2}W_{3}(x_{1},\xi )+\xi^{-d/2}W_{3}(x_{2},\xi^{-1})\right]
\label{eq:305}
\ee
where
\be
W_{3}(x_{1},\xi)=\int_{0}^{1}dy
\frac{y^{d/2-1}}{\left[(\xi +y)(1+\xi+y(1-y))\right]^{d/2}}
e^{-\frac{1}{2}x_{1}^{2}\tilde{g}_{0}(\xi ,y)}
\nonumber
\ee
\be
\xi=\frac{t_{2}}{t_{1}}
\nonumber
\ee
\be
x_{1}^{2}=r^{2}/(4 t_{1})
\nonumber
\ee
and
\be
\tilde{g}_{0}(\xi ,y)=\frac{4}{1+\xi +y(1-y)} ~~~.
\nonumber
\ee
Finally we need
\be
G_{4}^{(0)}(1112)=-\omega 2^{d}S_{2}(1)S_{2}(2) \xi^{d/4-1/2}
\nonumber
\ee
\be
\times\left[3W_{1}(x_{1},\xi )+W_{2}(x_{1},\xi )\right]
\label{eq:310}
\ee
\be
W_{1}(x_{1},\xi)=\int_{0}^{1}dy y^{d/2-1}
\left[\frac{2}{(1 +y)(1+3\xi+y(3-\xi )-2y^{2})}\right]^{d/2}
\nonumber
\ee
\be
\times e^{-\frac{1}{2}x_{1}^{2}\tilde{g}_{2}(\xi ,y)}
\ee
\be
W_{2}(x_{1},\xi )=\int_{0}^{\xi}dy y^{d/2-1}
\left[\frac{2}{(1 +y)^{2}(1+3\xi-2y)}\right]^{d/2}
e^{-\frac{1}{2}x_{1}^{2}\tilde{g}_{2}(\xi ,y)}
\nonumber
\ee
\be
\tilde{g}_{1}(\xi ,y)=\frac{6}{1+3\xi -2y)}
\nonumber
\ee
and
\be
\tilde{g}_{2}(\xi ,y)=\frac{4(1+\xi)}{1+3\xi +y(3-\xi )-2y^{2})}  ~~~.
\nonumber
\ee
Note the check on Eqs.(\ref{eq:305}) and (\ref{eq:310}) that they
reduce to Eq.(\ref{eq:303}) as $2\rightarrow 1$.
Pulling all of these results together we have
the ${\cal O}(1)$ corrections to the order-parameter correlation function
\be
C_{0,0}^{(1)}(12)=\frac{\omega 2^{d}\gamma^{3}}{3\pi}\Biggl[
-2L_{d}f(3-2f^{2})
\nonumber
\ee
\be
+\xi^{d/4}\left[3W_{1}(x_{1},\xi )+W_{2}(x_{1},\xi )\right]
\nonumber
\ee
\be
+\xi^{-d/4}\left[3W_{1}(x_{2},\xi^{-1})+W_{2}(x_{2},\xi^{-1} )\right]
\nonumber
\ee
\be
-3\xi^{d/2} f W_{3}(x_{1},\xi )-3\xi^{-d/2} f W_{3}(x_{2},\xi^{-1})\Biggr]
{}~~~.
\ee
There are several limits of interest.  First consider the onsite
unequal-time $t_{1}\gg t_{2}$ limit which gives the nonequilibrium
exponent $\lambda$.  In this limit $f$ is small and has the form
for large $\xi$ given by Eq.(\ref{eq:178}) which, in our notation
here, reads:
\be
f=A_{f}^{(0)}\xi^{-d/4}
\ee
where $A_{f}^{(0)}=2^{d/2}$
at lowest order.  Similarly the zeroth-order order parameter correlation
function has the form
\be
C^{(0)}(0,t_{1},t_{2})=\frac{2}{\pi}\psi_{0}^{2} A_{f}^{(0)}\xi^{-d/4}
{}~~~.
\ee
Second-order contributions to $f$ change the exponent $\lambda$ from $d/2$
to the expression given by Eq.(\ref{eq:14}).  Do the new first-order
terms give contributions which change $\lambda$ at first order?
To answer this question we need to set $r=0$ and take $\xi$ large
in $C_{0,0}^{(1)}(12)$.  The main results we need for large $\xi$ are:
\be
W_{1}(0,\xi )=\xi^{-d/2}q_{1}(d)
\nonumber
\ee
\be
W_{2}(0,\xi )=\xi^{-d/2}q_{0}(d)
\nonumber
\ee
\be
W_{1}(0,\xi^{-1} )=q_{2}(d)
\nonumber
\ee
\be
W_{2}(0,\xi^{-1} )=\frac{4}{d}d^{d/2} \xi^{-d/2}
\nonumber
\ee
\be
W_{3}(0,\xi )=\frac{2}{d}\xi^{-d}
\nonumber
\ee
\be
W_{3}(0,\xi^{-1} )=\left[ln ~\xi +q_{3}(d)\right]
\nonumber
\ee
with the dimensionality dependent quantities defined by
\be
q_{0}(d)=\left(\frac{2}{3}\right)^{d/2}\frac{\Gamma^{2}(d/2)}{\Gamma (d)}
\nonumber
\ee
\be
q_{1}(d)=\int_{0}^{1}dy~y^{d/2-1}\left[\frac{2}{(1+y)(3-y)}\right]^{d/2}
=2^{d/2}K_{d}
\nonumber
\ee
\be
q_{2}(d)=\int_{0}^{1}dy~y^{d/2-1}
\left[\frac{2}{(1+y)(1+3y-2y^{2})}\right]^{d/2} ~~~.
\nonumber
\ee
We see then that these contributions do not contribute terms to the
exponent $\lambda$ at first order.  One has only a contribution to the
amplitude
\be
C^{(1)}(0,t_{1},t_{2})=\frac{\omega 2^{d}}{3\pi}
\xi^{-d/4}\left[-6f_{0}L_{d}+3q_{1}(d)+q_{0}(d)+3q_{2}(d)\right]
{}.
\nonumber
\ee

For equal times $t_{1}=t_{2}=t, \xi=1,$ the expression
for $C_{00}^{(1)}(12)$  simplifies
significantly.  All of the W's share the same integrand except for the
$\tilde{g}$'s multiplying $x^{2}$ in the argument of the exponential.
After considerable manipulation we have
\be
C_{0,0}^{(1)}(12)=\frac{\omega 2^{d}\gamma^{3}}{3\pi}
\int_{0}^{1}
dy\frac{y^{d/2-1}}{\left[(2-y)(1+y)^{2}\right]^{d/2}}
\label{eq:350}
\ee
\be
\times\left[-2e^{-\frac{1}{2}x^{2}}(3-2f^{2})
+2e^{-\frac{1}{2}x^{2}\tilde{g}_{1}(1,y)}
+6(1-f)e^{-\frac{1}{2}x^{2}\tilde{g}_{0}(1,y)}\right]
{}.
\nonumber
\ee
For large x one has, since $\tilde{g}_{0}(1,y)> 1$, $\tilde{g}_{1}(1,y)> 1$, that only the
fully-contracted four-point cumulant contributes and
\be
C_{0,0}^{(1)}(12)=\frac{\omega 2^{d}}{3\pi}(-6L_{d}) f(12)
\ee
which does not give a first-order contribution to the exponent
$\nu$.

The last point to be discussed in the ${\cal O}(1)$ evaluation of
the order-parameter correlation function is that
one must be careful about the
behavior  at short-scaled distances at equal times.  Let us
write
\be
C_{0,0}^{(1)}(12)=\gamma^{3}\Delta (x)
\ee
where $\Delta (x)$ can be read off from Eq.(\ref{eq:350}).  Then
for small $x$
we see that $\Delta (x)$ goes to zero as $x^{2}$ while
$\gamma^{3}$ blows up as $1/x^{3}$.  Such singularities are
unphysical and indicate that we are expanding about a singular
order-parameter interfacial profile in this regime.  It is
just these singularities which give rise to Porod's law.
One possible resummation is
\be
\F (x)=\frac{2}{\pi}sin^{-1}
\left[f(x)-\frac{\pi}{2}\frac{\Delta (x)}{q_{0}}\right]
\nonumber
\ee
\be
-\frac{\pi}{2q_{0}}\Delta (x)\frac{1}{\sqrt{1-f(x)^{2}+4q_{0}/\pi}}
\ee
and a reasonable choice for $q_{0}$ is
$q_{0}^{2}=\Delta (\bar{x})$,
where $\bar{x}$ is some renormalization point chosen such that the
scaling function is smooth.

We can then conclude that the first-order correction to the order
parameter correlation function do not lead to corrections to the
exponents $\lambda$ and $\nu$.  Thus to ${\cal O}(1)$ $G_{2}(12)$
and $C(12)$ share the same OJK exponents.

\subsection{Terms of ${\cal O}(2)$}

Turning to the more involved case of the terms of ${\cal O}(2)$
contributions to $C(12)$, we find four types of terms.  There
are the terms in $G_{2}(12)$ which are of ${\cal O}(2)$ which must
be included in the contribution given by Eq.(\ref{eq:D25})  and which have
already been evaluated.  Inserting these results for $f$ into Eq.(\ref{eq:D25})
we find, since $f$ is small in the regime controlled by $\nu$ and
$\lambda$ that the order-parameter correlation function picks up the
same corrections leading to $\lambda_{m}$ and $\nu_{m}$.

Let us turn now to
the other three contributions:

i).  One has terms of ${\cal O}(2)$ from Eq.(\ref{eq:297})
where the leading order $G_{4}^{(1)}$ contributions are replaced by
their next-order corrections $G_{4}^{(2)}$.

ii).  The leading $G_{6}$ contribution in Eq.(\ref{eq:309}) is
of ${\cal O}(2)$ and contributes to $C^{(2)}(12)$.

iii).  The term $\left(G_{{\cal H}}^{(4)}\right)^{2}$ in
Eq.(\ref{eq:309}) is also of ${\cal O}(2)$.

Just as in Eq.(\ref{eq:311}) the factors of $k_{i}$ in Eq.(\ref{eq:309})
just lead to higher-order subscripts in the
multiplicative factors of $C_{n,\ell}^{(0)}(12)$.  Thus in
$C^{(2)}(12)$ there are contributions of the form
\be
\int~dx_{1}~\sigma (x_{1})\int~dx_{2}~\sigma (x_{2})
\int ~\frac{dk_{1}}{2\pi}\frac{dk_{2}}{2\pi}
e^{-ik_{1}x_{1}}e^{-ik_{2}x_{2}}
\nonumber
\ee
\be
\times\frac{20}{6!}
(ik_{1})^{3}(ik_{2})^{3} G_{6}(111222)
\nonumber
\ee
\be
=C_{3,3}^{(0)}(12) \frac{20}{6!} G_{6}(111222)
\nonumber
\ee
and
\be
\int~dx_{1}~\sigma (x_{1})\int~dx_{2}~\sigma (x_{2})
\int ~\frac{dk_{1}}{2\pi}\frac{dk_{2}}{2\pi}
e^{-ik_{1}x_{1}}e^{-ik_{2}x_{2}}
\nonumber
\ee
\be
\times\frac{1}{2}
\left(\frac{6(ik_{1})^{2}(ik_{2})^{2}}{4!}\right)^{2}
G_{4}^{2}(1122)
\nonumber
\ee
\be
=C_{4,4}^{(0)}(12) \frac{36}{2(4!)^{2}} G_{4}^{2}(1122) ~~~.
\nonumber
\ee
While there are a great many terms to be analyzed from all three sets
of contributions, none appear to give corrections to the exponents
$\lambda$ and $\nu$.  Thus it appears to ${\cal O}(2)$ that
$\lambda = \lambda_{m}$
and
$\nu = \nu_{m}$ ~.

\section{Equation of Motion Considerations}

In the last section we determined the order-parameter correlation function by
relating it to the auxiliary field correlation function.  It is also
instructive to consider using equation of motion methods for
determining the order-parameter scaling function as in
Ref.(\onlinecite{TUG}).  Let us return to
the equation of motion for the order
parameter given by Eq.(\ref{eq:45}).
We will
limit the discussion for simplicity to the case of equal times
where the equation of motion for the order-parameter correlation
function can be written
\be
\left(\frac{\partial}{\partial t}-2\nabla_{R}^{2}\right)
C(\vec{R},t)=-2<\sigma_{2}(1)(\nabla m(1))^{2}\sigma (2)>
\nonumber
\ee
\be
\equiv -2K(12) ~~~.
\ee
Assuming we have a scaling solution,
$C(\vec{R},t)=F(x) $,
we easily find that the equation of motion takes the form
\be
\vec{x}\cdot\nabla F(x)+\nabla^{2}F(x)=L^{2}K(12) ~~~.
\ee
We can then focus on $K(12)$.  At leading order it is easy
to show
that $K(12)$ can be written in terms of the two-point
probability distribution as
\be
K(12)=<(\nabla m(1))^{2}>C_{20}(12)+\bar{K}^{(0)}(12)
\ee
where $C_{20}(12)$ can be evaluated using Eq.(\ref{eq:D25}) in
Eq.(\ref{eq:299}):
\be
C_{20}^{(0)}(12)=-\frac{2}{\pi \left(S_{2}^{(0)}(1)\right)^{2}}f(x)\gamma (x) ~~~.
\ee
At lowest order we can also evaluate
\be
\bar{K}^{(0)}(12)=\Biggl[\nabla_{i}^{(3)}\nabla_{i}^{(4)}
\int~dx_{1}\int~dx_{2}~\sigma_{2} (x_{1})\sigma (x_{2})
\nonumber
\ee
\be
\times\frac{\delta^{2}}{\delta h(3)\delta h(4)}
P_{h}(x_{1}x_{2},12)\Biggr]|_{3=4=1,h=0}~~~.
\nonumber
\ee
\be
=\Bigg[\nabla_{i}^{(3)}\nabla_{i}^{(4)}\Big[
G_{2}^{(0)}(13)G_{2}^{(0)}(14)C_{40}(12)
\nonumber
\ee
\be
+G_{2}^{(0)}(23)G_{2}^{(0)}(24)C_{22}(12)
\nonumber
\ee
\be
+(G_{2}^{(0)}(13)G_{2}^{(0)}(24)+G_{2}^{(0)}(23)G_{2}^{(0)}(14))C_{32}(12)
\Big]\Bigg]|_{3=4=1}.
\nonumber
\ee
Since
\be
\left(\nabla_{i}^{(3)}G_{2}^{(0)}(13)\right)|_{3=1}=
\left(\nabla_{i}^{(4)}G_{2}^{(0)}(14)\right)|_{4=1}=0 ~~~,
\nonumber
\ee
we find
\be
\bar{K}^{(0)}(12)=\left(\nabla_{i}^{(1)}G_{2}^{(0)}(12)\right)^{2}
C_{22}(12) ~~~.
\ee
Putting these results together, using Eq.(\ref{eq:302}),
\be
f(x)\gamma (x)=tan[\frac{\pi}{2}F(x)]
\ee
and
\be
\nabla_{x}F(x)=\frac{2}{\pi}\gamma (x)\nabla_{x}f(x) ~~~,
\ee
we obtain the scaling equation
\be
\vec{x}\cdot\nabla F(x)+\nabla^{2}F(x)
\nonumber
\ee
\be
+tan[\frac{\pi}{2}F(x)]
\left[\frac{1}{\mu}-\frac{\pi}{2}(\nabla_{x}F(x))^{2}\right]=0
\label{eq:354}
\ee
where
\be
\frac{\pi}{2\mu}=\frac{L^{2}<(\nabla m)^{2}>_{0}}{S_{2}^{(0)}} ~~~.
\ee
Analyzing Eq.(\ref{eq:354}), however,
one finds
the remarkable result that there is an analytical solution given by
\be
F=\frac{2}{\pi}sin^{-1}\left[e^{-\frac{x^{2}}{2}}\right]
\ee
with $\mu $ given by the OJK result
$\mu =\frac{\pi}{2d}$ ~.
Thus the equation of motion method gives the same result as the
method of evaluating $C_{\psi}(12)$ directly.  This lends strong
support to the structure of the approach.

One interesting question concerns the work here compared to that
in TUG. If one makes the replacement
\be
\frac{1}{\mu}-\frac{\pi}{2}\left(\nabla F\right)^{2}
\rightarrow \frac{1}{\mu}
\ee
in Eq.(\ref{eq:354}) one obtains the basic equation in the TUG approach where
$\mu$ must then be determined as the solution to a nonlinear eigenvalue
problem.  Given the selected $\mu^{*}$ one then has the analytic results:
$\lambda =d-\frac{\pi}{4\mu^{*}}$,
and
$\nu=d-\frac{\pi}{2\mu^{*}}$
{}~.
In the TUG approach $\lambda$ and $\nu$ are not independent
($\nu =2\lambda -d$).
The  TUG results for $\lambda$ and $\nu$ as functions of $d$ are shown in
Tables I and II.  It seems clear that the TUG values for
$\lambda$ and $\nu$ are superior to those of the present second-order
theory.  In particular the TUG approach gives the known exact results
in one dimension.  Clearly the behavior of $\nu$ with $d$ is different in
the two approximations with TUG giving $\nu\rightarrow 0$ and
the current theory giving $\nu\rightarrow \infty$ as $d$ increases.
Thus the TUG results are in agreement with the speculation of Bray and
Humayun that large $d$ corresponds to the OJK  limit.

It should be kept in mind that while the TUG approach does well in
giving these exponents it is deficient in treating the smoothness
of the auxiliary field correlation function which enters into the
determination of defect dynamics.  The present theory gives good
results for the exponents-the corrections to OJK all appear to be in
the right direction and give the auxiliary field correlation functions
which are smooth.

\section{Conclusions}

The key accomplishment in this paper is to show how one can set up a
systematic calculation which allows for both nontrivial nonequilibrium
exponents, $\lambda$ and $\nu$, and results for the auxiliary field
correlation function which are  smooth enough to offer physical
results for defect structures.  The structure of the theory presented
here is appealing since the deviation from the OJK results can be handled
perturbatively and one can see and control the post-gaussian corrections.

The key question remaining is the uniqueness of the theory developed here.
While it seems that there is a degree of universality in this problem, the
answer to the question posed involves the robust  nature of this
universality.  This can be investigated by looking at those changes in the
equation of motion for the auxiliary field which may change scaling
results.  Thus one can try to find marginal variables which could lead
to exponents which depend on a parameter.  It seems reasonable
that the uniqueness of the particular realization of the theory presented
here can be investigated within its own structure.

It is clear that one can
introduce more sophisticated resummation methods than the direct method used
here.  These methods could be useful, for example, in establishing the
connection between the exponents governing the statistics of the auxiliary
field and those governing the statistics of the order-parameter field.
While these are equal at ${\cal O}(2)$, it is not at all obvious
that this holds at higher order.  It will be interesting to see if one
can formulate
alternatives to and resummations of the
$\phi_{p}$ expansion developed here.  For example
it seems to
desirable to find an expansion,  like in TUG, which matches onto the
exact solution for $d=1$.

In the second paper in this series the theory is extended to the
n-vector model.  It will turn out, at least at ${\cal O}(2)$, that
the $\phi$-expansion is related to a large-$n$ expansion.  The main
focus of this second paper will be to look at the determination of
defect spatial and velocity correlations.

\centerline{Acknowledgments}

I thank Dr. R. Wickham for useful comments.
This work was supported in part by the MRSEC Program of the National Science
Foundation under Award Number DMR-9400379.

\appendix

\section*{Solution for $S_{2}^{(0)}(t)$}

In this appendix we present the solution to the equation
\be
S_{2}^{(0)}(t)=exp\left(2\bar{\omega}\int_{t_{0}}^{t}
\frac{d\tau}{S_{2}^{(0)}(\tau)}\right)
\frac{S_{2}^{(0)}(t_{0})}{Q^{d/2}(t)}
\nonumber
\ee
where
\be
Q(t)=1+\alpha (t-t_{0}) ~~~.
\nonumber
\ee
Cross multiply by $Q^{d/2}(t)$ and take the time derivative to obtain
the rather simple equation
\be
\dot{S}_{2}^{(0)}(t)+\frac{d\alpha}{2}\frac{S_{2}^{(0)}(t)}{Q(t)}
=2\bar{\omega} ~~~.
\nonumber
\ee
Introducing the integrating factor
\be
S_{2}^{(0)}(t)=exp\left(-\frac{d}{2}\int_{t_{0}}^{t}d\tau
\frac{\alpha}{Q(\tau )}\right)
X(t)
\nonumber
\ee
where
\be
\dot{X}(t)=exp\left(\frac{d}{2}\int_{t_{0}}^{t}d\tau
\frac{\alpha}{Q(\tau )}\right)
2\bar{\omega} ~~~.
\nonumber
\ee
This is easily integrated to obtain
\be
X(t)=S_{2}^{(0)}(t_{0})
+2\bar{\omega}\int_{t_{0}}^{t}d\bar{t} ~
exp\left(\frac{d}{2}\int_{t_{0}}^{\bar{t}}d\tau \frac{\alpha}{Q(\tau )}\right)
{}~~~.
\nonumber
\ee
We can then do the integral
\be
\int_{t_{0}}^{t}d\tau \frac{\alpha}{Q(\tau )}
=\alpha \int_{t_{0}}^{t}d\tau \frac{d}{d\tau}
ln [1+\alpha (\tau -t_{0})]
\nonumber
\ee
\be
=ln (1+\alpha (t-t_{0}))
{}~~~.
\nonumber
\ee
This result is then inserted in the integrating factor and the
remaining integral over $\bar{t}$ can easily be carried out to obtain
the result given by Eq.(\ref{eq:187}).

%
%

\newpage

\begin{table}
\begin{tabular}{|c|c|c|c|c|}
\hline
& & & &\\
$dimension$ &  $theory $ & $ OJK  $&  $TUG $ & $ best$\\
& & & &\\
$1$      & $0.6268..$ &$0.5$  & 1       &  $1^{a}$  \\
$2$      & $1.1051..$ &$1$    & 1.2887  &  $1.246\pm 0.02^{b}$  \\
$3$      & $1.5824..$ &$1.5$  & 1.6726  &  $1.838\pm 0.2^{b}$  \\
$large $ & $ d/2 $    &$d/2$  & d/2     &  $d/2^{c}$  \\
 & & & & \\
\hline
\end{tabular}
\caption{Values of exponent $\lambda $ from the current theory, OJK and TUG.
a.   Exact, see ref.(\protect\onlinecite{2}),
b. numerical results from ref.(\protect\onlinecite{LM91}),
c.  Best guess.}
\end{table}

\begin{table}
\begin{tabular}{|c|c|c|c|c|}
\hline
& & & & \\
$dimension$ &  $theory $ &  $OJK  $&  $TUG $ &$\omega$\\
& & & &\\
$1$      & $1.1596..$       &$0$  & $ 1 $ & $0.4601..$ \\
$2$      & $1.2492..$       &$0$  & $ 0.5774.. $& $0.6877..$ \\
$3$      & $1.3732..$       &$0$  & $ 0.3452.. $ & $0.9067..$\\
$large $ & $ d(3-2\sqrt{2}) $ & $0 $ &$ 0 $ & $\frac{d}{2}(\sqrt{2}-1) $  \\
& & &  &\\
\hline
\end{tabular}
\caption{Values of exponent $\nu$ from the current theory, OJK and TUG.
The last column gives the values of the  quantity $\omega$.}
\end{table}


\begin{thebibliography}{99}
\bibitem{1/d}  There has been speculation, see Ref.\onlinecite{2},
that the OJK result is obtained in a systematic expansion in powers
of 1/d.  We do not find support for this conjecture here.
\bibitem{second}  The nature of the expansion is discussed below.
Convergence properties, if any, remain to be investigated.
\bibitem{TUG3}
In TUG
it was assumed that $m$ is governed by a gaussian probability distribution
and the variance is determined by demanding that the equation of motion
satisfied by the order parameter by satisfied on average
\be
<\Lambda (1) \psi (1)\psi (2)>=- <V'[\psi (1)]\psi (2)>  ~~~.
\nonumber
\ee
\bibitem{uniarg}  This is one of the arguments in favor of the
universality of the results obtained
\bibitem{causality}
One point to keep in
mind is the well known property of the MSR method that the
equal-time correlation functions with auxiliary
fields must vanish if one is to
statisfy causality:
$G_{Mm}(11)=0$.
\end{thebibliography}
\end{document}